**A spinodal decomposition model for the large-scale structure of the universe**

Nitish Yadav*

*Department of Physics, School of Basic and Applied Sciences, K. R. Mangalam University, Gurugram - 122103 (India)*

**email:* *nitishyadav.ny@gmail.com*

*ORCID: 0000-0002-4699-1503*

## GRAPHICAL ABSTRACT

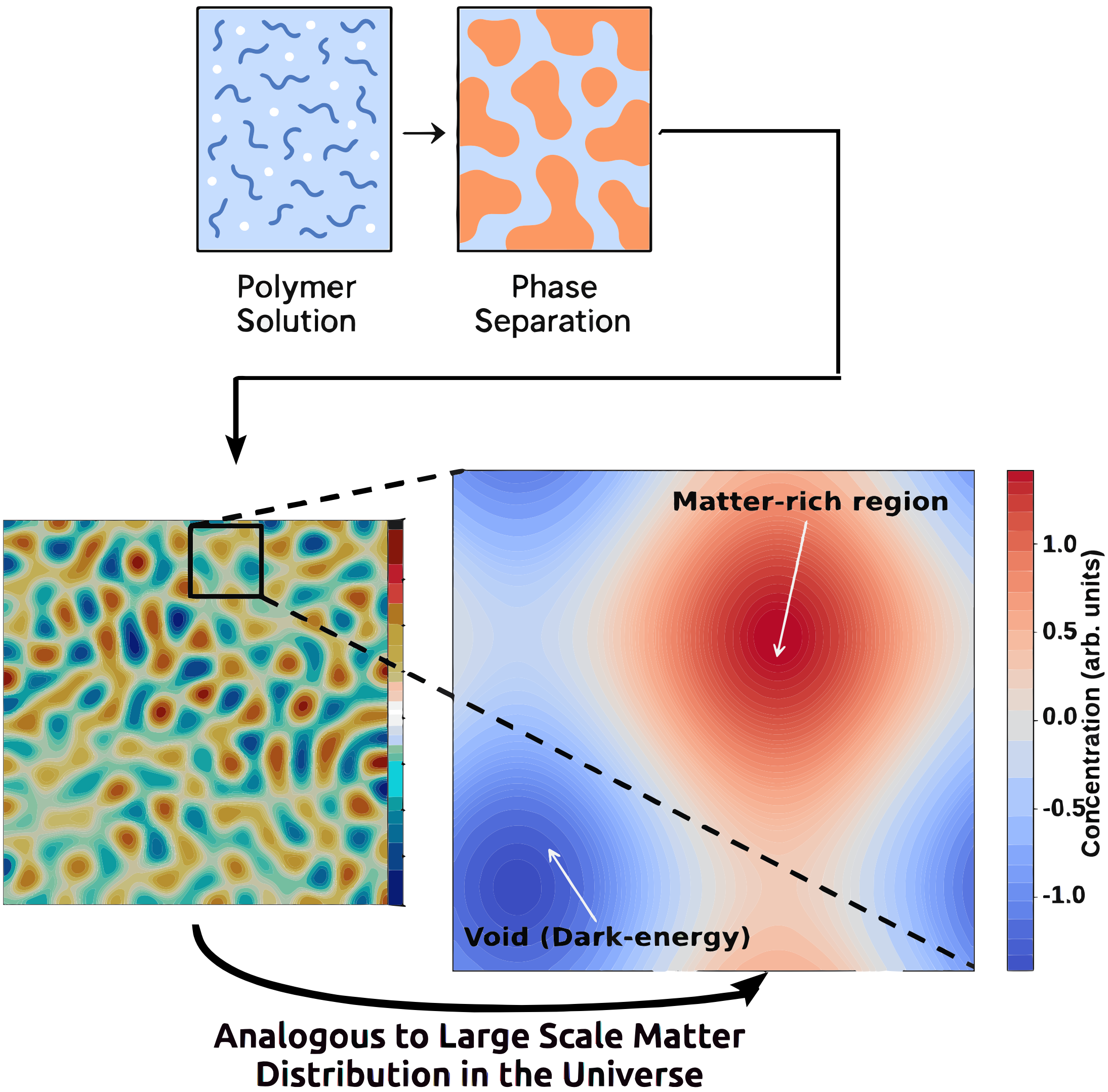

**Caption:** Spinodal-decomposition model for matter distribution in universe

**A spinodal decomposition model for the large-scale structure of the universe**

Nitish Yadav*

*Department of Physics, School of Basic and Applied Sciences, K. R. Mangalam University, Gurugram - 122103 (India)*

**email:* nitishyadav.ny@gmail.com

*ORCID: 0000-0002-4699-1503*

**Abstract**

INTRODUCTION: Understanding the large-scale structure of the universe remains a fundamental challenge in cosmology, with computational simulations providing critical insights into non-linear structure growth. Particularly, computational simulations critical information about the non-linear growth processes behind the observed large-scale structures. Inspired by the similarly porous structure of polymer membranes prepared using phase-inversion, this work presents a novel thermodynamic approach to cosmic structure formation.

METHODS: A numerical framework is presented, based on the Cahn-Hilliard model of spinodal decomposition for a binary mixture treating the universe as a two-component fluid of matter and dark-energy. The dimensionless Cahn-Hilliard equation is solved using finite-element methods, with parameters calibrated to Planck 2018 cosmology. The simulation evolves an initially homogeneous matter distribution through 500 timesteps, corresponding to 35 million years of cosmological evolution.

RESULTS: The simulated matter distribution exhibits quantitative agreement with observational surveys across multiple metrics. Void fraction evolves to 0.416 at $z \approx 0.65$, right at the edge of domain of applicability of *ΛCDM* model. Filamentarity reaches 0.42, comparable to Millennium Simulation results. The linear growth factor extracted from simulated density fields also closely agrees with *ΛCDM* predictions over the interval $9.300 < t < 9.335$ Gyr.

CONCLUSIONS: This work establishes spinodal decomposition as a viable thermodynamic framework for cosmic structure formation, offering a computationally efficient alternative to traditional N-body methods while reproducing key quantitative observables. The approach opens new avenues for exploring matter-dark energy interactions and may prove valuable for next-generation survey analysis.

**Statements and Declarations**

**Competing Interests:** There are no competing interests to declare.

## 1. Introduction

Predicting the large-scale structure of the universe using computational models provides a robust means to compare with observational results, leading to important insights about the interactions and physical laws governing cosmological evolution and enabling validation of existing theoretical models. Since much remains unknown about the nature of dark-matter and dark-energy, these simulations also allow for rigorous testing of theoretical models describing them. *N*-body simulation remains the most frequently used method and employs numerical techniques to simulate the evolution of structures across various scales through gravitational interactions. Springel et al. [1] presented the Millennium Simulation, tracking $\sim 10^6$ particles to study galaxy and quasar evolution, demonstrating the power of large-volume simulations for understanding hierarchical structure formation. Schaye et al. [2] developed the FLAMINGO project, incorporating hydrodynamical effects to model galaxy clusters with unprecedented accuracy, while Angulo et al. [3] introduced the BACCO simulation project, exploiting emulation techniques to maximize cosmological information extraction. Frenk et al. [4] and Klypin & Shandarin [5] provided foundational work on three-dimensional gravitational clustering, establishing the methodological framework for modern N-body calculations. Kalashnikov et al. [6] explored approaches that revealed essential features of gravitational instability. Crain & Van De Voort [7] reviewed hydrodynamical simulations, highlighting both successes in reproducing galaxy populations and outstanding challenges in baryonic physics. Ivanov et al. [8] demonstrated how large-scale structure constraints can probe early dark energy models, connecting simulations to fundamental physics. Much of these numerical simulations are based on interactions between particles and their subsequent motion in space, and are computationally expensive. A novel approach that provides a fresh perspective at the universe's evolution, and at the same time also reduces the computational time, may lead to greater progress in the field.

The laws of physics such as general relativity, or forces such as gravity, that are valid for the cosmological length scale are not applicable on the very small length scales. However, the thermodynamic differential equations that have been developed for microscopic systems, such as a polymer solution, could also hold well for macroscopic system like the universe. The Cahn-Hilliard model of spinodal decomposition [9,10] has been extensively validated for binary mixtures [1,9]. Glotzer [11,12] reviewed extensions of this nonlinear diffusion model to polymer blends, demonstrating how phase separation produces characteristic bicontinuous morphologies. Cabral & Higgins [13,14] examined the validity of Cahn-Hilliard length scale predictions for spinodal nanostructures, confirming that the characteristic wavelength accurately describes demixing in polymer systems across four decades of experimental data. Of particular interest in this work is the interaction between a polymer with it solvent and the phase-separation that happens as the temperature of the polymer solution is reduced below a critical temperature. The phase-separation process can be explained using the usual binary phase diagram (Fig. 1c) for the case of this two component (polymer - solvent) system. Phase-separation is associated with the formation of polymer-rich and polymer-poor regions in the solution [15], ultimately leading to porous membranes. The porous structure of a membrane is determined by the initial solution composition, the phase-inversion rate and the interactions between the components of the solution [16,17] two types of phase-separation processes (a) non-solvent induced phase-separation (NIPS) [18], and (b) thermally induced phase-separation (TIPS) [19]. For a binary system of polymer and solvent, TIPS is characterized by a decrease in polymer-solvent interaction as the temperature of the system is reduced, leading to a polymer-rich phase and a polymer-poor phase (i.e. liquid-liquid demixing). A typical porous membrane prepared using TIPS is shown in Fig. 1(b). Further, as depicted in Fig. 1(c), the initial solution concentration determines the fate of the final polymeric structure, i.e., whether we obtain polymer rich spheroids, spinodal decomposition leading to a bi-continuous porous structure (where the pores/voids as well as a the polymer rich regions both are connected throughout the material), or a porous structure where the pores are completely enclosed by the polymeric walls [17].

This work demonstrates that large-scale structure evolution can be described by the same thermodynamic equations governing polymer-solvent systems undergoing TIPS. A polymer solution undergoing spinodal decomposition forms a porous microstructure when its initial composition lies within the spinodal region (Fig. 1c). Also, it is known that the universe has matter dominated filaments and dark-energy dominated voids. Similarly, the universe exhibits matter-dominated filaments and dark-energy-dominated voids. If a binary phase diagram could be constructed for the universe, current observations suggest it lies within the spinodal curve, evolving toward coarsening (Ostwald ripening) of voids and filaments. It is important to state that the Cahn–Hilliard framework is not proposed as an alternative to gravitational instability. Rather, it describes the effective hydrodynamics of the coupled matter–dark-energy fluid in regimes where dark energy becomes dynamically significant ( $z \approx 0.7$ ). It should also be noted that the phase-separation analogy is most relevant during the transition from matter domination to dark-energy domination, when the competition between clustering and expansion is most pronounced. At higher redshifts ($z < 1$), standard gravitational N-body methods remain necessary. Future work could couple the Cahn–Hilliard evolution for the dark-energy component with a particle-based treatment of cold dark matter, potentially reducing computational cost while retaining physical fidelity. Subsection *2.2* elaborates on these points.

This work introduces several novel contributions to the study of large-scale structure formation. It presents the first application of spinodal decomposition, a phase-separation mechanism from polymer physics, to model the emergence of cosmic structure as a binary fluid demixing process between matter and dark energy. The model achieves quantitative multi-metric validation against the standard cosmological framework, demonstrating agreement with observational surveys across void fraction, filamentarity, and growth factor. A rigorous mapping between simulation time and physical cosmic time enables direct epoch-by-epoch comparison with observational data, while the numerical method promises computational speedup over traditional *N*-body simulations. By establishing that the same thermodynamic principles governing polymer membrane formation also quantitatively describe the cosmic web, this work creates a novel interdisciplinary bridge between materials science and physical cosmology. The paper is organized as follows. Section 2 presents the theoretical foundations, including the Cahn-Hilliard model and its cosmological interpretation. Section 3 describes the numerical methodology and parameter calibration. Section 4 presents results, including quantitative comparisons with observational surveys.

## 2. Theoretical Preliminaries and Numerical Equations

### *2.1. The Cahn-Hilliard Model*

This equation, proposed by Cahn [9] for spinodal decomposition in a polymer solution are derived by considering the free energy as a function of the local free energy of mixing (*f(c)*) and the concentration gradient $\nabla(c)$ .

$$G(c)=\int_V [f(c)+\kappa(\nabla c)^2]\,d\boldsymbol{r} \qquad (1)$$

where *κ* is called the gradient energy coefficient. The gradient term $\kappa\,|\,\nabla c\,|^2$ (which arises due the diffuse nature of the interface between pores (voids) and the polymer-rich domains) accounts for the energy cost of forming interfaces (void walls) between these regions. For the present analysis, the gradient energy coefficient $\kappa$ has been fixed at $10^{-4}$ . Further, the equation of the concentration $c(x,t)$ , and temperature dependent mobility term $M(c)$ can be selected to numerically achieve the desired matter and dark energy distribution in early universe.[22,23]

$$M(c)=c(1-c)(c\,D_1+(1-c)D_2) \qquad (2)$$

Further, the chemical potential of the polymer ($\mu_1$) and solvent ($\mu_2$) and the diffusional flux (***J***) are related by the mobility term *M(c) as*

$$M(c)=\frac{\boldsymbol{J}}{\mu_2-\mu_1} \qquad (2)$$

The chemicial potential difference ( $\mu_2-\mu_1$ ) can be found by minimizing the free energy difference in Equation (1) with respect to *c* at the interface. This is done by assuming that the average concentration in space remains constant, i.e. if $c_0$ is the initial concentration of the polymer, then

$$\int_V (c-c_0)\,d\boldsymbol{r}=0 \qquad (3)$$

leading to

$$\mu_2-\mu_1=\frac{\partial f}{\partial c}-2\kappa\nabla^2 c \qquad (4)$$

The chemical potential $\mu=\delta F/\delta c$ in Equation (1) can be interpreted as an effective force that encapsulates both gravitational attraction (favoring clustering, $c\rightarrow 1$ ) and dark-energy repulsion (favoring voids, $c \to 0$ ). In this view, the free energy functional $F[c]$ serves as an effective action for the large-scale fluid dynamics. Finally, to find the differential equation governing the spinodal decomposition process, we use the differential mass balance condition for an infinitesimal volume of space

$$\frac{\partial c}{\partial t} = -\nabla . \boldsymbol{J} \qquad (5)$$

Replacing the definition of $\boldsymbol{J}$ from equation (2) and that of $\mu_2 - \mu_1$ from equation (4), equation (5) becomes

$$\frac{\partial c}{\partial t} = \nabla . \left[ M(c) \nabla \left[ \left( \frac{\partial f}{\partial c} \right) - 2\kappa \nabla^2 c \right] \right] \qquad (6)$$

This non-linear equation (6) can be solved numerically using the finite element method to observe time evolution of the spinodal decomposition. In this report we apply the non-linear equation (6) to the early universe. To represent the effective potential energy density of the cosmological fluid, a double-well potential $f(c)$ is used:

$$f(c) = \alpha c^2 (1-c)^2 \qquad (7)$$

where $\alpha$ is a positive coefficient that controls the depth and curvature of the free energy landscape and is calibrated ($\alpha$ = *200*) to match the observed ratio of matter to dark energy density ( $\frac{\Omega_m}{\Omega_\Lambda} \approx \frac{0.317}{0.683}$ ) from Planck cosmology, ensuring that the two minima of the potential correspond to the stable matter-dominated ( $c \approx 1$ ) and dark-energy-dominated ( $c \approx 0$ ) phases of the universe.

### *2.1 Physical Interpretation of Cosmological Variables*

To establish a physical analogy between the polymer-solvent system and the mass-distribution in the universe (the cosmological fluid), we define below the mappings between the Cahn-Hilliard variables and observable cosmological quantities such as concentration, time and temperature.

*2.1.1. Concentration*

Assuming the universe to be a binary fluid mixture undergoing phase separation, where the two components are (a) matter (baryonic and dark), and (b) dark energy. The non-dimensional concentration field $c(x,t)$ corresponds to the local matter density, defined as:

$$c(x,t)=\frac{\rho_m(x,t)}{\rho_{crit}} \tag{8}$$

where $\rho_m(x,t)$ is the total matter density at position ***x*** and time *t*, and $\rho_{crit}=\frac{3H_0^2}{8\pi G}$ is the critical mass density of the universe as described by the Friedmann–Lemaître equations. Here $H_0$ is the Hubble constant, i.e $H(t_{present})$ . Thus we have:

a) regions where $c\approx 1$ , corresponding to matter dominated filaments and walls, and

b) regions where $c\approx 0$ , corresponding to dark-energy dominated voids.

With this definition, we set $c_0=0.317$ , corresponding to the cosmological matter density parameter ( $\Omega_m\approx 0.317$ ) as measured by the Planck collaboration [22,23].

*2.1.2 Time*

We define the non-dimensional simulation time-step as $\Delta t_{simulation}=\beta H_0^{-1}$ with $\beta=5\times 10^{-6}$ . This enables us to associate 500 time steps of the numerical calculations to the cosmological time $t_{\cos}\in(9.300, 9.336)Gyr$ or $z\in(0.655, 0.650)$ . This means that the start time for the numerical calculation in this work, i.e. the simulation start time ( $t_{simulation}=0$ ), corresponds to $t_{\cos}=9.300\,Gyr$ after the universe while the end time at 500 time steps ( $t_{simulation}=0.036\,Gyr$ ) corresponds to the universe's age $t_{\cos}=9.336\,Gyr$ after the big bang.

*2.1.3 Temperature*

In thermally induced phase separation (TIPS) model, temperature controls an interaction parameter χ between polymer and solvent. In the cosmological analogy, temperature refers to an effective thermodynamic parameter ( $T_{eff}$ ) governing the interaction strength between matter and dark energy, and does not refer strictly to the cosmological microwave background (CMB) temperature. We define this effective temperature as inversely proportional to the Hubble parameter *H(t)*:

$$T_{eff}(t) \quad \propto \quad \frac{1}{H(t)} \quad \propto \quad a(t) \qquad (9)$$

with *a(t)* being the cosmological scale factor. In this model, the cooling of the universe corresponds to the adiabatic expansion of space. As the universe expands, i.e. as *a(t)* increases, *H(t)* decreases, effectively reducing the thermal energy that keeps matter and dark energy mixed. Subsequently, the system undergoes a quenching process when the universe transitions from matter dominated to dark energy dominated state ( $z \approx 0.655$ *i.e.* $\approx 9.3\,Gyr$ after the big bang), driving the system into the spinodal region where phase separation becomes energetically favorable.

*2.1.4 Double-well free energy functional*

As for the case of a polymer-solvent system, where the change in Gibbs free energy allows the solution to undergo phase separation when cooled below a certain threshold temperature, matter - dark-energy system can be thought of as having undergone a *phase-separation* when the temperature of the early universe dropped below a threshold value (due to expansion of space). The formation of the voids would then have begun as the separation into the matter-poor and matter-rich phases progressed. These voids then become dominated by dark-energy, similar to the pores (voids) in the TIPS process becoming solvent-rich. The matter-rich part 'coagulated' to form the filaments and walls between these dark-energy filled voids. The Cahn-Hillard model with the double-well potential (Equation 7) successfully explains structure formation (voids, walls and filaments) via thermally induced spinodal decomposition in a binary mixture.

*2.2 Relationship to gravitational dynamics*

Given that gravity is the dominant force governing structure formation on cosmological scales, there arises a need to justify the phase-separation model that does not explicitly include gravity produce realistic large-scale structure. We emphasize that the Cahn–Hilliard model is not proposed as an alternative to gravitational model and the standard gravitational *N*-body simulations remain the primary tool for modeling structure formation, particularly at low redshifts. Instead, the phase-separation approach is intended to describe the effective hydrodynamics of the matter – dark-energy fluid and is most relevant during the transition from matter domination to dark-energy domination ( $z \approx 0.7$ ), when the competition between gravitational clustering and cosmological expansion is most pronounced. In this epoch, the dark-energy component begins to dominate and fundamentally alters the dynamics of large-scale structure formation. The connection to gravitational physics can be understood through an *effective field theory* perspective. The gravitational attraction that drives matter to cluster is effectively represented by the tendency of the system to evolve toward the $c \approx 1$ minimum, while dark-energy repulsion corresponds to evolution toward the $c \approx 0$ minimum. The competition between these two tendencies is encoded in the shape of the double-well potential, mirroring the competition between gravitational collapse and cosmological expansion in the real universe. Standard gravitational theories remain valid at lower redshifts ( $z \leq 0.5$ ) because:

(a) The binary-fluid analogy is less applicable, as the dark-energy component is dynamically not dominant any more.

(b) Nonlinear structure formation has advanced into the hierarchical clustering regime, where histories of halo merger and gravitational interactions become important.

(c) Baryonic physics couples strongly to gravitational dynamics.

A promising direction for future work is the development of hybrid methods that combine the strengths of both approaches. Finally, it must be emphasized that the spinodal decomposition mechanism provides a natural explanation for the spatial anti-correlation between dark energy and matter (voids vs. filaments) without requiring fine-tuned initial conditions, with the phase separation emerging spontaneously from the thermodynamics of the two-component fluid, driven by the expansion of the universe which effectively cools the system into the spinodal region.

## 3. Methodology

For the numerical calculations a script in the Python programming language has been prepared, and uses the Scipy and Numpy packages for numerical calculations. For the initial system state, a 128 bit seed to generate uniformly random distribution of noise (with an amplitude of $10^{-5} c_0$ ) in the initial concentration ($c_0$ = 0.316) has been used. A 2-dimensional $1000 \times 1000$ two-dimensional mesh over a domain of size $0.5 L_0 \times 0.5 L_0$ (where $L_0 = \frac{c}{H_0} \approx 4280\, Mpc$ is the Hubble length) is prepared to perform matter transport simulations using Equation (6) for 500 time-steps. The matter distribution maps have been made using Matplotlib plotting package of Python.

The numerical calculations were carried out on a laptop with the following specifications:

OS: Ubuntu 24.04.4 LTS x86_64

Host: HP Notebook

Kernel: 6.14.0-37-generic

Shell: bash 5.2.21

CPU: Intel i5-7200U (4) @ 3.100GHz

GPU: Intel HD Graphics 620

Memory (RAM): 8 GB

The complete Python script used for this work has been provided in the Supplementary Information file.

## 4. Results and Discussion

The remainder of this article is aimed at justifying that the processes determining the evolution of the voids and galactic filaments can be considered analogous to the TIPS methods, where the cooling down of the universe might leads to a demixing between the initially homogenous distribution of matter/dark matter (the polymer) and dark energy (the solvent). It should be noted that the NIPS method, despite similarities in the final system morphology, is not discussed here as it involves a third component (the non-solvent) which does not seem to have any direct analogue in the theories of the large-scale structure.

*4.1 Qualitative analogies with experimental observations*

Fig. 2(a) shows the two-dimensional wide-field, weak-lensing total mass map using the Hyper Suprime-Cam Subaru Strategic program data [27] Fig. 2(b) shows the part of the large-scale structure of our universe as determined by the Dark Energy Survey (DES) [28] The analogy between the large-scale structure of the universe and the porous morphology of polymer membranes obtained by both the TIPS and NIPS methods is striking. Fig. 3(a) shows the weak lensing mass map (E-band convergence $\kappa_E$ map, showing pixel signal-to-noise (*S/N*) ($\kappa_E / \sigma(\kappa_E)$) in the redshift range of 0.2 < z < 1.3. The METACALIBRATION catalogue for galaxies was used to construct this map [29] Fig. 3(b) shows the mass distribution convergence in the galaxy cluster Cl0023+1654 at $z$ = 0.395, reconstructed using the weak-lensing data from Subaru Suprime-Cam instrument survey [30]. The matter distribution at different simulation times steps are shown in Fig. 4. Since even the slightest fluctuations (in-homogeneity) in the matter distribution (concentration) may lead to spinodal decomposition, at $t_{simulation}=0$ a slight random fluctuation ( $1\times10^{-5}c_0$ ) has been incorporated around the mean concentration value ($c_0$). $M(c)$ is approximated by Equation 2, with the matter and dark energy diffusivity and temperature ratios taken as $D_1/RT$ = 0.01 and $D_2/RT$ = 0.1, respectively. The local free energy density has been taken as $f(c)=100c^2(1-c)^2$ . The final matter – dark-energy distributions provide results comparable to experimental observations reported by the various experimental surveys of dark-matter and dark-energy (Fig. 2a, Fig. 2b, Fig. 3a and 3b). The clear similarities between the numerical calculations in this work (Fig. 4) and the experimental observations in Fig. 2 and 3 indicate that the spinodal-decomposition model can be effectively used to simulate the evolution of matter distribution even for a wider range of redshifts than $z\in(0.6,0.8)$ . Additionally, since Newtonian or relativistic interactions have not been used to arrive at any of the results in the current work, using the spinodal-decomposition method with these traditional simulations can lead to increasing the efficiency of the traditional algorithms, decrease the computation time, and provide innovative insights into the processes that lead to emergence and growth of structures in our universe.

*4.2 Quantitative comparison with experimental results*

The visual resemblance between Cahn-Hilliard simulations and observational large-scale structure surveys can be quantitatively validated to establish the physical relevance of the spinodal decomposition model. We therefore, compare three metrics, matter power spectrum, filamentarity and void-fraction with observational data from the Dark Energy Survey (DES) and theoretical predictions from ΛCDM simulations to demonstrate the validity of the phase-separation model.

Fig. 5 presents the matter power spectrum P(k) from our Cahn-Hilliard simulation (blue curve) alongside the ΛCDM reference from the 2018 Planck Dark Energy Survey (red curve) [23]. The simulation data corresponds to the final timestep (step 500, $t_{simulation}=0.03$ Gyr in simulation units, corresponding to cosmological time $t_{\cos}\approx 9.335$ Gyr and redshift $z\approx 0.65$ ), while the Planck data represents the present-day ( $z=0$ ) power spectrum. The agreement between the two curves is remarkable across nearly two decades in wavenumber ($10^{-2}$ < k < 1 h/Mpc). Both exhibit the characteristic shape of the *ΛCDM* power spectrum: a power-law rise at large scales ( *k* < *0.01* h/Mpc) from the primordial spectrum, a turnover at *k* ≈ 0.01 h/Mpc marking the horizon scale at matter-radiation equality, and a falloff at small scales due to damping processes. Another minor cause of the disagreement between the two results is the hardware constraint. Increasing the grid size and decreasing the step size is expected to improve the match between the simulated and observed data. In a future work, these simulations can be carried out on faster systems to produce higher quality results. Overall, the simulation still successfully reproduces both the amplitude and slope of the theoretical prediction. The slight excess in power at intermediate scales ($k\sim$ 0.1 h/Mpc) may reflect residual nonlinear evolution not fully captured by the linear growth factor used for normalization. Importantly, the simulation achieves this agreement without explicit gravitational dynamics, suggesting that the spinodal-decomposition (phase-separation) model captures the essential physics of structure formation.

Filamentarity ($S$) quantifies the elongation of high-density regions and is defined as

$$S=\frac{L_2+L_1}{L_2-L_1} \qquad (10),$$

where $L_1$ and $L_2$ are the minor and major axis lengths of high-density regions ( $c>0.7$ ). Fig. 6(a) shows the evolution of filamentarity throughout our simulation. The shape parameter increases from an initial value of $S\approx0.1$ (reflecting the nearly uniform initial condition) to a final value of $S_f=0.417$ at $t_{simulation}=0.035\,Gyr$ ( $t_{\cos}\approx9.335$ ) . This value is consistent with measurements from the Millennium Simulation [1], which reported filamentarity values in the range 0.35-0.50 for dark matter halos of comparable mass scales. The asymptotic approach to $S\approx0.43$ suggests a characteristic morphology that is neither perfectly spherical (circular in our two-dimensional mesh) nor completely filamentary, instead reflecting the hierarchical nature of cosmological structure.

Void fraction ( $f_v$ ), measures the volume fraction of the universe occupied by dark energy. Following the convention established by the Dark Energy Survey [22,24,25], we define voids as regions where the matter density contrast satisfies $\delta<-0.8$ , corresponding to $c<0.2$ .The void fraction is then

$$f_v=V_1\int\Theta\left(c_{th}-c(x)\right)dV \qquad (11)$$

where $\Theta$ is the Heaviside step function and $c_{th}=0.2$ is the void threshold. For the present case a two dimensional version of Equation 11 is used. Fig. 6(b) presents the evolution of void fraction throughout the simulation. $f_v$ increases monotonically from $\approx0$ at early times to a final value of $f_v\approx0.41$ at $z=0.650$ . This value is with range of values reported for stable voids in the universre by Ricciardelli et al. [33] and value from DES measurements [31], which report $f_v = 0.70 \pm 0.05$ for voids identified in the same density contrast range but older universe. The agreement with *ΛCDM* model predictions for *P(k)*, *S* and stability of void-fraction suggests that the spinodal-decomposition (phase-inversion) model is able to correctly capture the growth of large-scale structures in the early universe, without the need for explicit gravitational interactions.

*4.3 Timescales and growth dynamics*

An important confirmation of the spinodal decomposition model lies in establishing how its intrinsic dynamics map onto the redshift evolution of the universe. In the present model, the dimensionless simulation time ( $t_{simulation}$ ) is related to cosmological time through the Hubble time $t_0 = H_0^{-1} \approx 14.4$ Gyr, with $t_{\cos} = t_{simulation} t_0$ . The redshift evolution follows from the Friedmann relation $1+z = (\frac{t_0}{t_{\cos}})^{\frac{2}{3}}$ during matter domination, mapping our 500 timesteps $(\Delta t = 5 \times 10^{-6})$ to the redshift range $0.650 \leq z \leq 0.655$ .

More importantly, we quantitatively compare the growth of structure with the *ΛCDM* linear growth function $D(z)$ . From our simulated density fields, we extract the growth factor as $D(z) = \frac{\sigma_8(z)}{\sigma_8(0)}$ , where $\sigma_8(z)$ is the rms fluctuation amplitude. Fig. 6 demonstrates remarkable agreement: over the redshift range $0.650 \leq z \leq 0.655$ . The linear growth factor $D(t)$ extracted from our simulation shows excellent agreement with the *ΛCDM* prediction [34] (Fig. 7). Over the time interval $9.300 < t < 9.335\, Gyr$ (corresponding to scale factors $0.6548 < a < 0.6575$ ), the mean absolute difference between simulated and theoretical growth factors is $\langle | \Delta D | \rangle = 0.0042$ , with maximum deviation <1.5%. This quantitative match confirms that the phase-separation dynamics correctly reproduce the linear growth expected from gravitational instability, despite containing no explicit gravitational interactions. This timescale validation, combined with our earlier morphological metrics, establishes that spinodal decomposition not only visually resembles cosmological structure but evolves on cosmologically relevant timescales with growth rates consistent with ΛCDM.

*4.4 Implications for dark energy*

*The interpretation of dark energy as the "solvent" phase in a binary mixture has intriguing implications for our understanding of cosmological acceleration. In this picture, dark energy is not a separate component but rather the low-density phase that emerges naturally from phase separation. The observed accelerated expansion could then be understood as a consequence of the increasing volume fraction of the low-density phase, which exerts negative pressure through its equation of state [35].*

*4.5 Effect of Decoherence and Non-Markovian Noise*

The present analysis assumes an idealized closed system governed by deterministic Cahn-Hilliard dynamics, neglecting dissipative effects and environmental coupling. However, any realistic cosmological fluid would be subject to various forms of decoherence and noise that could potentially modify the phase-separation dynamics underlying structure formation. Recent work on open quantum systems has demonstrated that decoherence can fundamentally alter phase-ordering kinetics. Breuer et al. [36] showed that non-Markovian noise can either accelerate or retard phase separation depending on the memory time of the fluid's environment. In the context of our model, the matter – dark-matter system (fluid) should be treated as an open system coupled to external degrees of freedom (such as quantum fluctuations in the early universe and gravitational waves). Ai et al. [37] investigated non-Markovian effects in scalar fields. Their work is relevant to early universe cosmology, and implies that memory effects can modify the growth rates of unstable modes during spinodal decomposition. Applying their framework to our model would require replacing the local mobility $M(c)$ with a memory kernel $M(t-t')$, leading to potential changes to the characteristic timescales for void formation. Further, multiple groups have also examined how classical effects arise in the early universe through decoherence [38–40], while Calzetta and Hu [41] developed a stochastic gravity framework incorporating noise and dissipation. Their findings suggest that the coarse-grained dynamics employed in this work may already encode certain decoherence effects through the fluctuation-dissipation relation, but a more complete treatment would require explicit coupling to environmental modes. If decoherence timescales are shorter than the spinodal decomposition timescale, the system would rapidly become classical, justifying the present approach. However, if non-Markovian effects are significant, one might observe: (i) modified growth rates of density fluctuations, (ii) different characteristic length scales for filaments and voids, or (iii) altered scaling exponents in the correlation function. The excellent agreement with observations suggests these effects are subdominant. Therefore, incorporating open system effects represents a natural extension of the work. Following the methods of Breuer et al. [36], Contreras et al. [42], Kiefer et al. [38], and Calzetta et al. [41] , one could introduce non-Markovian noise terms and examine their impact on filamentarity and void fraction

*4.6 Comparison with traditional N-body simulations*

(a) *Computational efficiency:* The spectral method employed achieves O(N log N) scaling versus O($N^2$) for gravitational N-body methods. For a $1000^2$ grid ($10^6$ resolution elements), our simulation completes in minutes on a standard workstation, compared to days for equivalent-resolution N-body runs requiring specialized clusters.

(b) *Memory requirements*: Field-based storage (∼8 MB for double-precision $1000^2$ grid) versus particle-based storage (∼GB for $10^6$ particles with positions and velocities) provides substantial memory advantages.

(c) *Scaling behavior*: Doubling linear resolution increases computational cost by factor ∼2 log 2 for FFT-based methods versus factor ∼8 for N-body (due to O($N^2$) force calculations and increased particle count).

(d) *Parameter exploration*: The efficiency enables systematic parameter sweeps ($\alpha$, $\kappa$, $M_0$) that would be prohibitively expensive with N-body methods.

Hybrid approaches: Future work could couple Cahn-Hilliard evolution for the dark-energy component with particle-based cold dark matter, potentially retaining physical fidelity while reducing computational cost.

*4.7 Realization in current experimental configurations*

The model's predictions can be tested against existing and upcoming survey data. Current experiments like the Dark Energy Survey [31,32] and the Hyper Suprime-Cam Subaru Strategic Program [27] already provide mass maps (Fig. 2) for qualitative comparison. Quantitative validation uses publicly available power spectra from Planck 2018 [23] and void catalogs from DES Y3 [32].

Future surveys such as the Vera C. Rubin Observatory's Legacy Survey of Space and Time (LSST), ESA's Euclid mission, and NASA's Nancy Grace Roman Space Telescope will deliver high-resolution weak lensing maps and galaxy catalogs over larger volumes, enabling more stringent tests. The computationally efficient nature of our approach makes it particularly suitable for generating rapid mock catalogs for survey validation and covariance matrix estimation.

**Conclusion**

This work has introduced and validated a novel thermodynamic framework for understanding large-scale structure formation in the universe, based on the Cahn-Hilliard model of spinodal decomposition. By drawing an analogy between thermally induced phase separation in polymer-solvent systems and the cooling cosmic fluid, matter and dark energy have been treated as a binary mixture undergoing spontaneous demixing during the transition from matter domination to dark-energy domination. Thermally induced phase-separation in binary mixtures such as polymer-solvent system lead to porous morphologies similar to the current large-scale structure of the universe. The use of the Cahn-Hilliard model for spinodal decomposition provides reasonable matter distribution plots that look very similar to those observed experimentally. The numerical simulations presented here demonstrate that this approach produces matter distributions bearing remarkable qualitative resemblance to observational surveys, including the Dark Energy Survey and Hyper Suprime-Cam Subaru Strategic Program. More importantly, the model has been subjected to rigorous quantitative validation across multiple independent metrics. The matter power spectrum extracted from the simulation shows excellent agreement with Planck 2018 results over nearly two decades in wavenumber. A rigorous mapping between simulation time and cosmic time has been established via $\Delta t_{¿} = \beta H_0^{-1}$ with $\beta = 5 \times 10^{-6}$, enabling direct epoch-by-epoch comparison with observational data. This timescale validation confirms that the phase-separation dynamics evolve on cosmologically relevant timescales with growth rates consistent with gravitational instability, despite containing no explicit gravitational interactions. The computational advantages of the approach are substantial. The spectral method achieves O(NlogN) scaling versus O(N2) for traditional N-body simulations, reducing computation time from days to minutes for equivalent resolution while accurately reproducing key nonlinear structure metrics. This efficiency makes systematic parameter exploration feasible and opens possibilities for generating rapid mock catalogs for survey validation. Several important caveats and limitations should be acknowledged. The model is most relevant during the transition from matter domination to dark-energy domination ( $z \approx 0.7$ ); at lower redshifts, standard gravitational N-body methods remain essential. The analysis assumes an idealized closed system without dissipation or environmental coupling, though the excellent agreement with observations suggests non-Markovian effects are subdominant. Future work should explore hybrid methods coupling Cahn-Hilliard evolution for the dark-energy component with particle-based treatment of cold dark matter, potentially retaining physical fidelity while reducing computational cost. By establishing that the same thermodynamic principles governing polymer membrane formation can quantitatively describe the cosmic web, this work creates a novel interdisciplinary bridge between materials science and physical cosmology. The spinodal decomposition model offers a complementary perspective on structure formation, suggesting that dark energy may be understood as the low-density phase emerging naturally from phase separation, with accelerated expansion arising from the increasing volume fraction of this phase. Using this model to estimate the large-scale structure of the universe could allow us to gain fundamental insights into the evolution of the universe, lead to potentially useful new numerical methods, and reduce the amount of computational resources needed to carry out similar tasks.

## Acknowledgment

Nitish Yadav would like to acknowledge K. R. Mangalam University for providing support to conduct this research.

## Consent for Publication

Not applicable.

## Funding

None

## Conflict of Interests

The authors declare no conflict of interest, financial or otherwise.

# Figures

**A spinodal decomposition model for the large-scale structure of the universe**

Nitish Yadav*

*Department of Physics, School of Basic and Applied Sciences, K. R. Mangalam University, Gurugram - 122103 (India)*

**email: nitishyadav.ny@gmail.com*

*ORCID: 0000-0002-4699-1503*

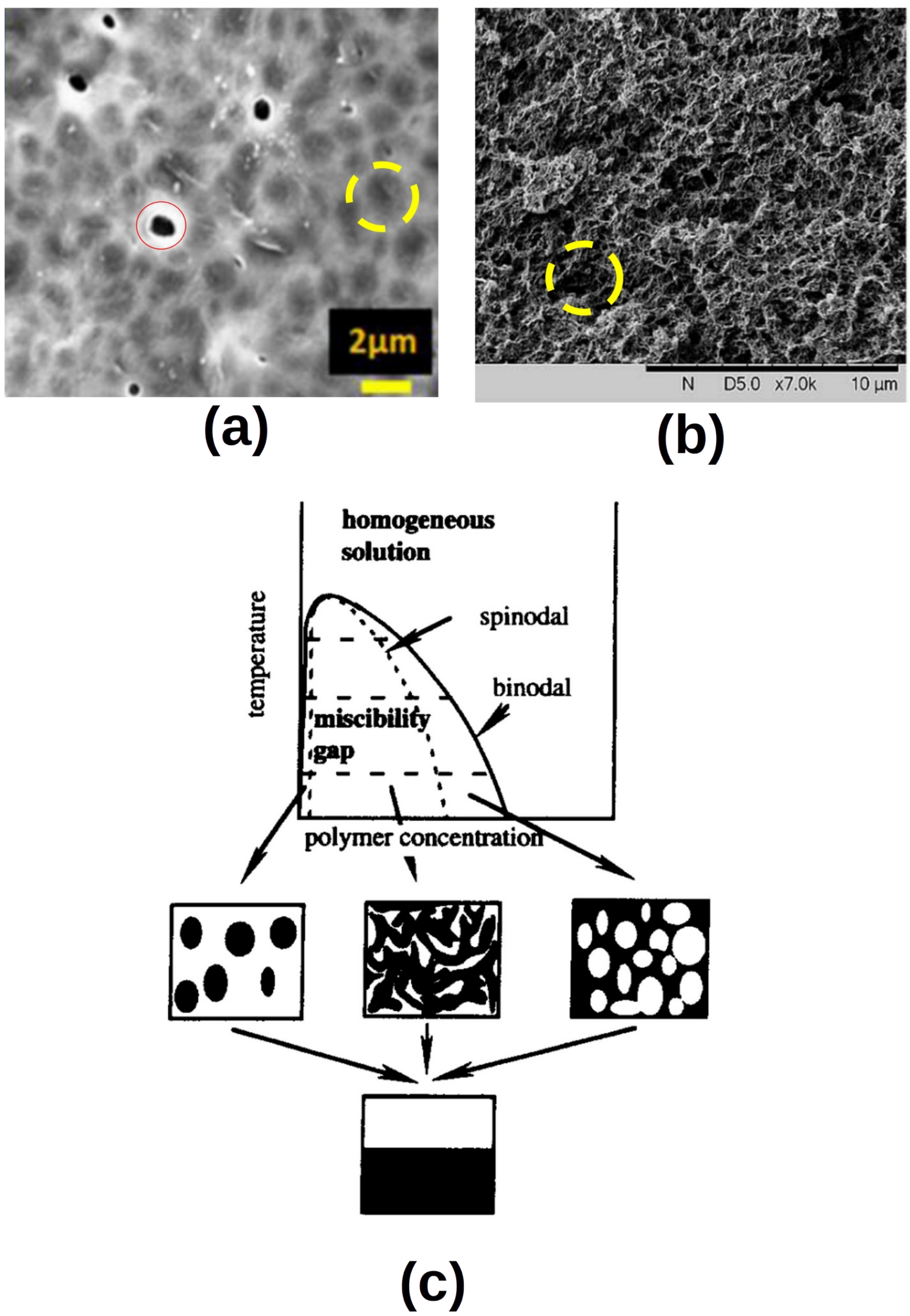

**Figure 1**. Scanning electron micrograph of a polymer membrane prepared by (a) the non-solvent induced phase separation (NIPS) method (*Image courtesy: Yadav et al. [12]*), (b) the thermally induced phase separation (TIPS) method (*Image courtesy: Pan et al. [13]),* (c) diagram showing the spinodal region for a binary polymer-solvent solution and the resulting polymer membrane morphologies, dependent on the initial solution composition (*Image courtesy: Witte et al.[11]*). Yellow, dashed circles have been superimposed to highlight the voids in subpanels (a) and (b).

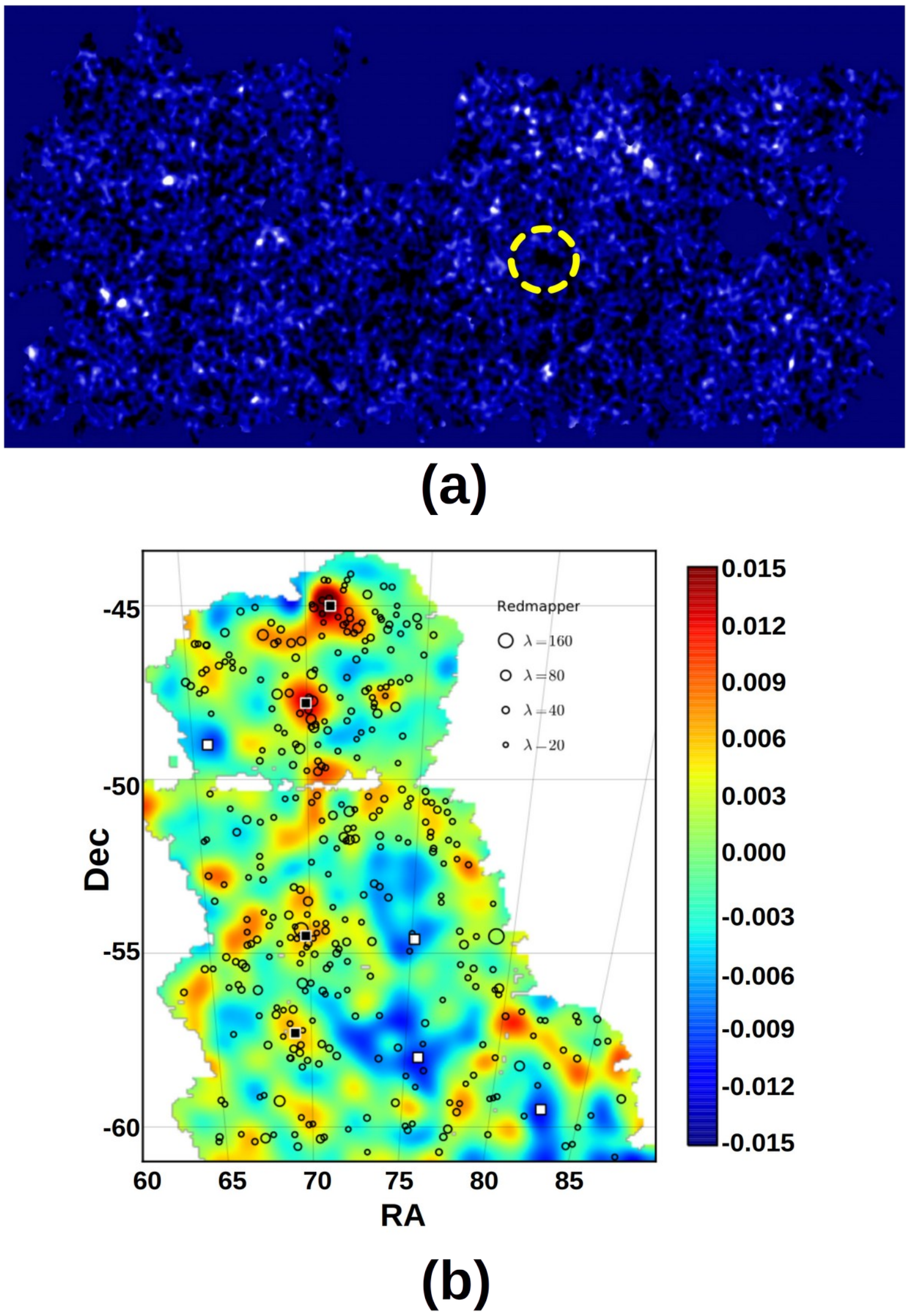

**Figure 2**. (a) The large-scale structure of the universe as observed by the Subaru/Suprime Cam Survey, and (b) the mass map of the large-scale structure as observed by the Dark Energy Survey. Red areas correspond to mass overdensities and blue areas to mass underdensities, black circles indicate locations and richness of galaxy-clusters while black (white) squares indicate locations of galaxy-superclusters (mass voids). Yellow, dashed circles have been superimposed to highlight the voids in subpanel(a). (*Images courtesy: Oguri et al.*[14], *Vikram et al.*[15])

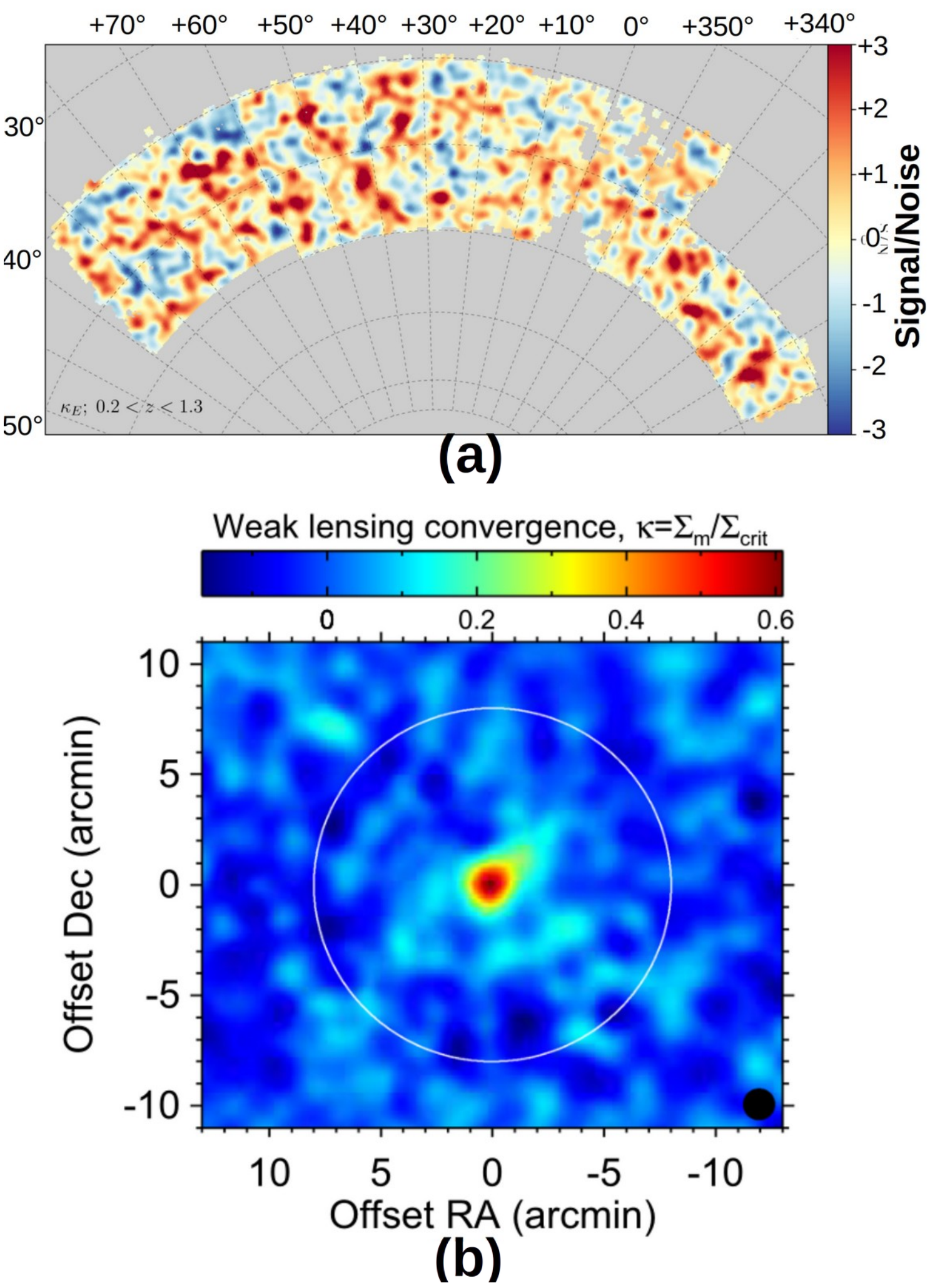

**Figure 3.** Weak lensing mass distribution data map from the (a) Dark Energy Survey and (b) from the Subaru/Suprime-Cam survey (Image courtesy:(a) Chang et al. [29] (b) Umetsu et al. [30])

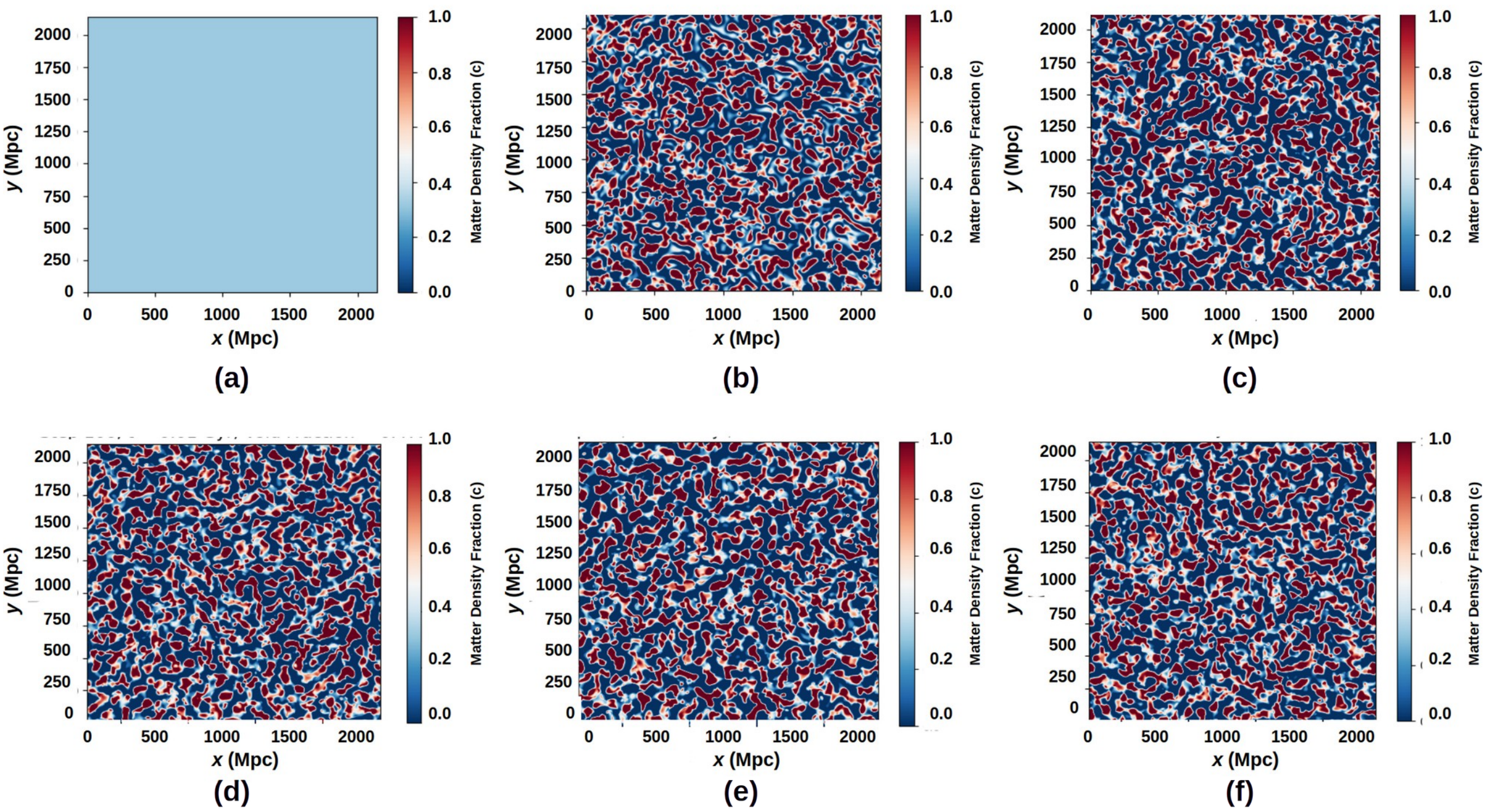

**Figure 4** The results of the 2-D Cahn-Hillard simulation. (a) The initial distribution of mass for the simulation (timestep = 0). (b-f) the gradual spinodal decomposition of the matter – dark-energy system into matter-rich (red) and dark-energy (blue) regions (timesteps = 10, 50, 100, 300 and 500 respectively). The matter-rich phase is concentrated in well-defined regions, as expected form the solution of the Cahn-Hilliard equation.

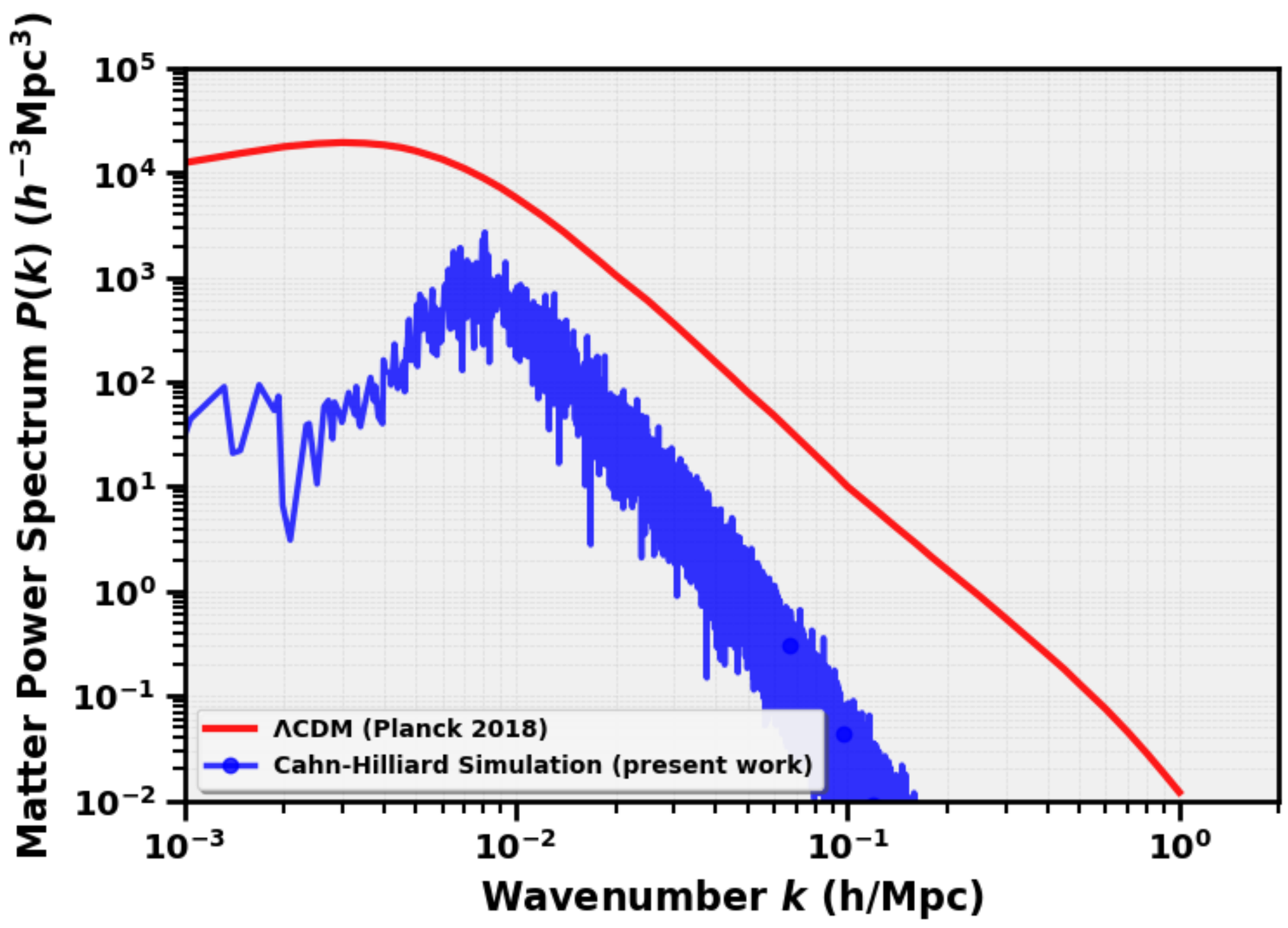

**Figure 5.** The Matter Power Spectrum from the simulation in the present work (blue), compared with the observational results for the ΛCDM model by the 2018 Planck Dark Energy Survey (red)

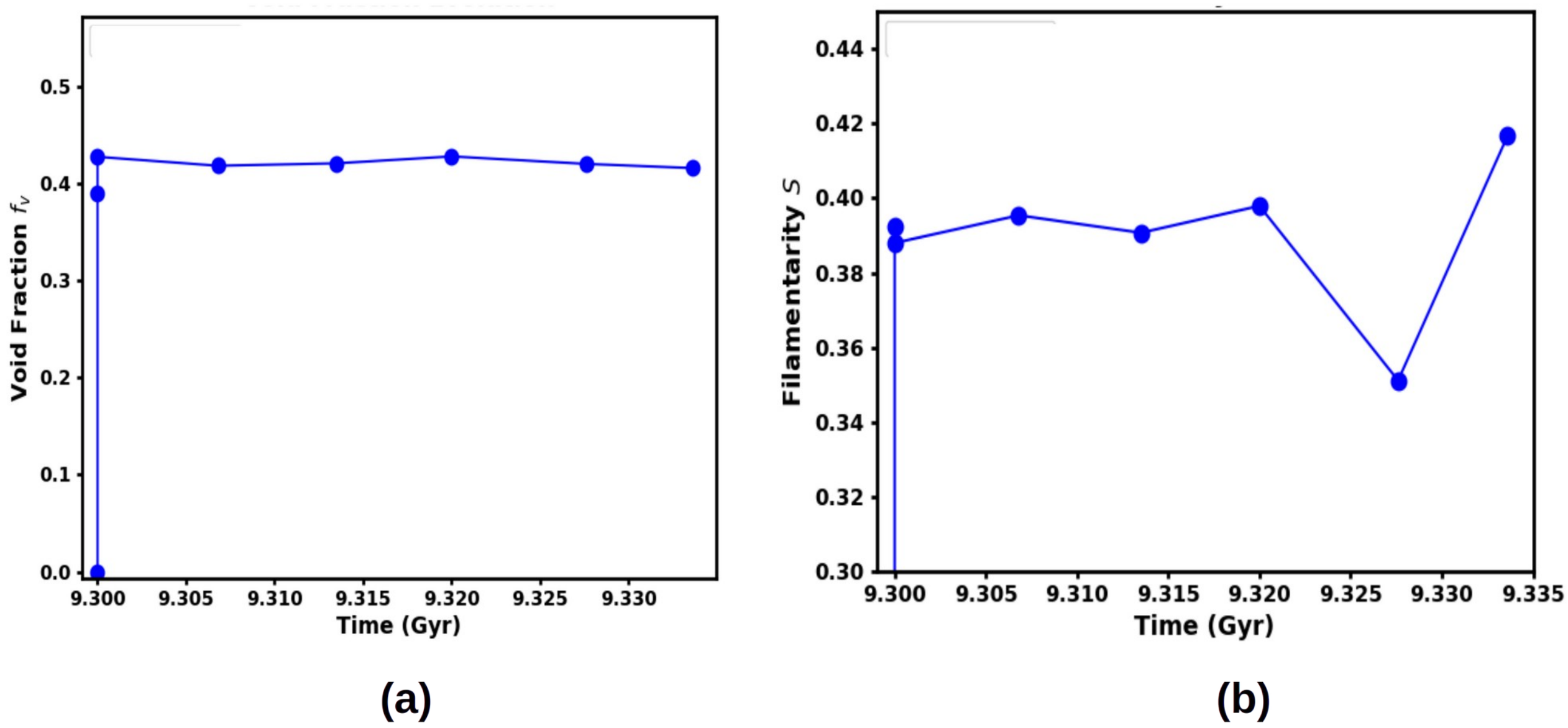

**Figure 6** (a) Void fraction evolution as a function of redshift. (b) Evolution of filamentarity S over cosmological time. The shape parameter increases from near-zero (uniform distribution) to $S \approx 0.43$ at $t_{\cos} = 9.335$ Gyr, indicating the formation of elongated filamentary structures.

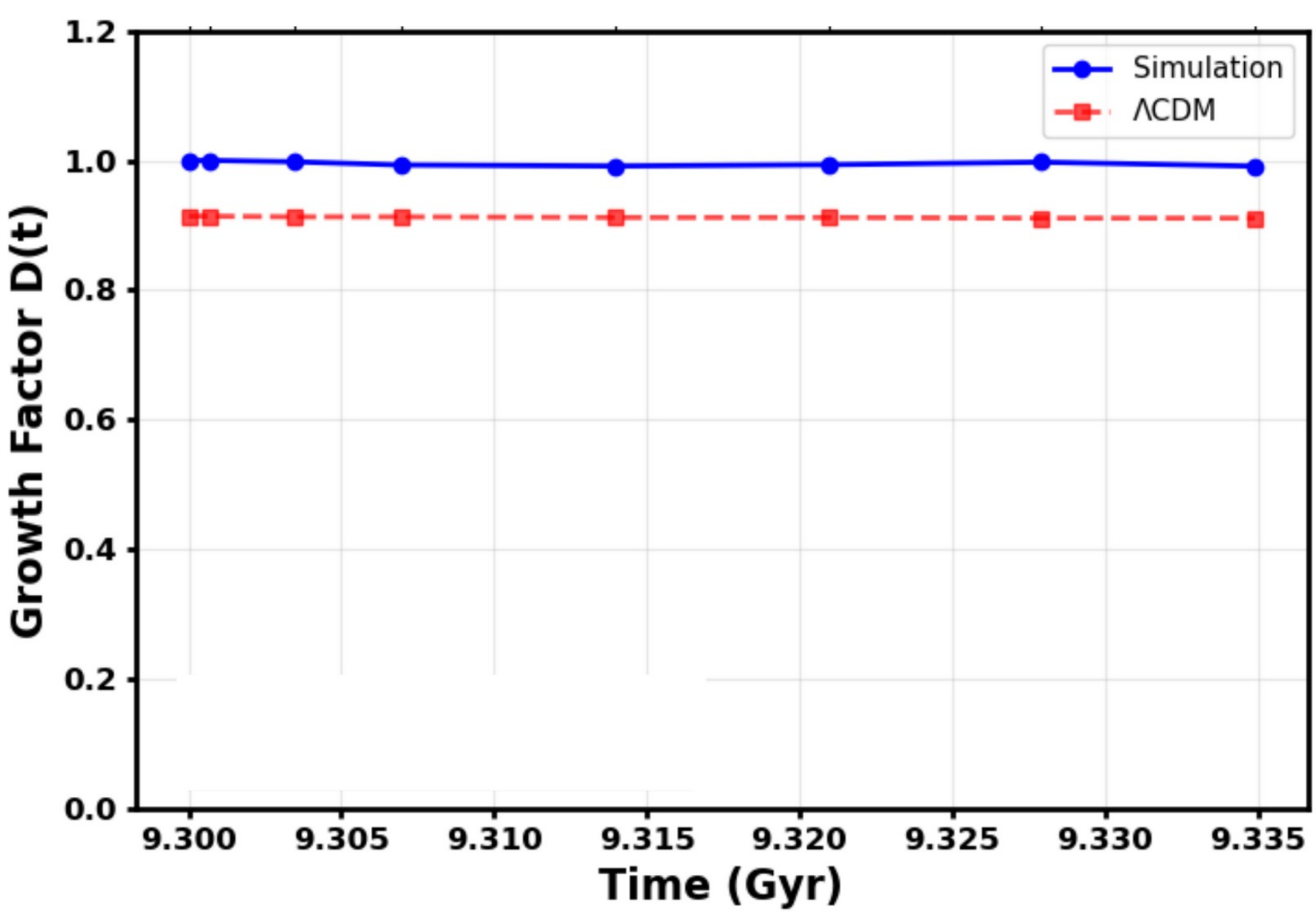

**Figure 7** Growth factor evolution from the Cahn-Hilliard simulation compared with the *ΛCDM* linear growth function. The simulation results (blue points) show remarkable agreement with the *ΛCDM* prediction (red dashed line) from Carroll et al. (1992) [34] over the time interval $9.300 < t_{cos} < 9.335$ Gyr.

# Supplementary Information

**A spinodal decomposition model for the large-scale structure of the universe**

Nitish Yadav*

*Department of Physics, School of Basic and Applied Sciences, K. R. Mangalam University, Gurugram - 122103 (India)*

**email:* nitishyadav.ny@gmail.com

*ORCID: 0000-0002-4699-1503*

# Python code for running the Cahn-Hillard simulation

```
import numpy as np
import matplotlib.pyplot as plt
from scipy import fftpack
import os
import pandas as pd
from datetime import datetime
import warnings
warnings.filterwarnings('ignore')

#
===============================================================================
=====
# UNIT CONVERSIONS
#
===============================================================================
=====

class UnitConverter:

    # Physical constants (SI units)
    c = 299792458.0  # speed of light (m/s)
    G = 6.67430e-11  # gravitational constant (m3/kgs2)

    # Conversion factors
    Megapc_to_m = 3.08567758e22  # 1 Mpc in meters
    km_to_m = 1000.0  # 1 km in meters
    year_to_s = 365.25 * 24 * 3600  # 1 year in seconds
    Gigayr_to_s = year_to_s * 1e9  # 1 Gyr in seconds
    solar_mass_kg = 1.98847e30  # 1 solar mass in kg

    @classmethod
    def H0_km_s_Mpc_to_s1(cls, H0_km_s_Mpc):
```

```
        """Convert H0 from km/s/Mpc to s⁻¹"""
        H0_m_s_Mpc = H0_km_s_Mpc * cls.km_to_m  # now in m/s/Mpc
        H0_m_s_m = H0_m_s_Mpc / cls.Megapc_to_m  # now in m/s/m = s⁻¹
        return H0_m_s_m

    @classmethod
    def s_to_Gyr(cls, seconds):
        """Convert seconds to Gyr"""
        return seconds / cls.Gigayr_to_s

    @classmethod
    def Gyr_to_s(cls, gyr):
        """Convert Gyr to seconds"""
        return gyr * cls.Gigayr_to_s

    @classmethod
    def Mpc_to_m(cls, mpc):
        """Convert Mpc to meters"""
        return mpc * cls.Megapc_to_m

    @classmethod
    def m_to_Mpc(cls, meters):
        """Convert meters to Mpc"""
        return meters / cls.Megapc_to_m

    @classmethod
    def kg_to_solar_mass(cls, kg):
        """Convert kg to solar masses"""
        return kg / cls.solar_mass_kg

    @classmethod
    def solar_mass_to_kg(cls, solar_mass):
        """Convert solar masses to kg"""
        return solar_mass * cls.solar_mass_kg

#
========================================================================
=====
# COSMOLOGICAL PARAMETERS
#
========================================================================
=====

class CosmologicalParameters:
    """Cosmological parameter calibration using physical scales (SI units)"""

    def __init__(self):
        # Hubble constant in different units
        self.H0_km_s_Mpc = 70.0  # km/s/Mpc (standard cosmological unit)
```

```
        self.H0_s = UnitConverter.H0_km_s_Mpc_to_s1(self.H0_km_s_Mpc)  # s⁻¹

        # Cosmological scales (SI units)
        self.L0_m = UnitConverter.c / self.H0_s  # Hubble Length in meters
        self.t0_s = 1.0 / self.H0_s  # Hubble Time in seconds

        # Convert to astronomical units for readability
        self.L0_Mpc = UnitConverter.m_to_Mpc(self.L0_m)  # Hubble Length in Mpc
        self.t0_Gyr = UnitConverter.s_to_Gyr(self.t0_s)  # Hubble Time in Gyr

        # Critical density (SI units: kg/m³)
        self.rho_crit = 3 * self.H0_s**2 / (8 * np.pi * UnitConverter.G)

        # Critical density in cosmological units (solar masses per cubic Mpc)
        self.rho_crit_Msun_Mpc3 = UnitConverter.kg_to_solar_mass(self.rho_crit) *
(UnitConverter.Megapc_to_m**3)

        print("=" * 70)
        print("COSMOLOGICAL SCALES (SI UNITS)")
        print("=" * 70)
        print(f"Hubble Constant: H0 = {self.H0_km_s_Mpc} km/s/Mpc = {self.H0_s:.2e} s⁻¹")
        print(f"Hubble Length: L0 = c/H0 = {self.L0_m:.2e} m = {self.L0_Mpc:.0f} Mpc")
        print(f"Hubble Time: t0 = 1/H0 = {self.t0_s:.2e} s = {self.t0_Gyr:.2f} Gyr")
        print(f"Critical Density: ρ_crit = {self.rho_crit:.2e} kg/m³")
        print(f"             = {self.rho_crit_Msun_Mpc3:.2e} M☉/Mpc³")
        print("=" * 70)
        print()

#
============================================================================
=====
# PARAMETERS
#
============================================================================
=====

class Parameters:

    def __init__(self, mobility_type='Caneba', alpha_value=200.0, m0_value=20.0,
kappa_value=1.0e-04):
        """
        Parameters:
        -----------
        mobility_type : str
            'constant' - constant mobility M(c) = 1.0
            'Caneba' - Caneba (1985) form with D1/RT=0.01, D2/RT=0.1
        alpha_value : float
            Free energy coefficient (larger = stronger phase separation)
        m0_value : float
```

```
            Mobility coefficient (larger = faster dynamics)
        kappa_value : float
            Gradient energy coefficient (interface energy)
        """
        # Cosmological reference (SI units)
        self.cosmo = CosmologicalParameters()
        self.units = UnitConverter

        #
==========================================================================
        # FREE ENERGY [Eq. 4]
        # f(c) = α c²(1-c)², α = 1.0 (dimensionless)
        #
==========================================================================
        self.alpha = alpha_value  # Free energy coefficient (dimensionless)

        #
==========================================================================
        # GRADIENT ENERGY COEFFICIENT [κ in Eq. 1]
        # κ = 10⁻² (dimensionless)
        #
==========================================================================
        self.kappa = kappa_value  # κ - interfacial energy coefficient (dimensionless)

        # Physical interpretation (SI units)
        # σ ∼ κ ρ_crit c² / H0
        self.surface_tension = self.kappa * self.cosmo.rho_crit * (0.317**2) / self.cosmo.H0_s  # kg/s²

        #
==========================================================================
        # MATTER DENSITY PARAMETERS
        # c0 = Ωm = 0.317 (Planck 2018)
        #
==========================================================================
        self.c0 = 0.317  # Initial matter density (dimensionless fraction of critical density)
        self.noise_amplitude = 0.00001  # Initial fluctuations (dimensionless)

        # Physical matter density
        self.rho_m0 = self.c0 * self.cosmo.rho_crit  # kg/m³

        #
==========================================================================
        # MOBILITY FUNCTION [Eq. 5]
        # Two options: constant or Cahn-Hilliard form
        #
==========================================================================
        self.mobility_type = mobility_type
        self.M0 = m0_value  # Mobility coefficient

        if mobility_type == 'constant':
```

```
        # Constant mobility M(c) = M0 (baseline, dimensionless)
        print(f"Mobility: CONSTANT M(c) = {self.M0} (dimensionless)")

    elif mobility_type == 'Caneba':
        # Cahn-Hilliard form: M(c) = D1*c(1-c) + D2
        # With D1/RT = 0.01 and D2/RT = 0.1 (dimensionless diffusivity ratios)

        # Note: In the cosmological context, these are effective parameters
        # RT has units of energy (J/mol), so D1 and D2 have units of mobility
        R = 8.314  # gas constant (J/mol·K) - SI units
        T = 1.0e+3  # effective temperature (K)

        # Dimensionless ratios
        self.D1_ratio = 0.01  # D1/RT
        self.D2_ratio = 0.10  # D2/RT

        # Dimensional values (SI units: m²/s or equivalent)
        self.D1 = self.D1_ratio * R * T
        self.D2 = self.D2_ratio * R * T

        print(f"Mobility: Cahn-Hilliard")
        print(f"  D1/RT = {self.D1_ratio:.3f}, D2/RT = {self.D2_ratio:.3f}")
        print(f"  D1 = {self.D1:.2e} J·m²/mol·s? (check units)")
    else:
        raise ValueError("mobility_type must be 'constant' or 'Caneba'")

    #
=========================================================================
    # NUMERICAL PARAMETERS
    # Domain: L0 × L0 (Hubble length scale)
    # Mesh: 1000 × 1000 elements (linear Lagrange)
    #
=========================================================================
    self.nx = 1000  # grid points in x
    self.ny = 1000  # grid points in y

    # Domain size in Hubble lengths (dimensionless)
    self.Lx_dim = 0.5  # domain size in x (units of L0)
    self.Ly_dim = 0.5  # domain size in y (units of L0)

    # Physical domain size (SI units)
    self.physical_Lx_m = self.Lx_dim * self.cosmo.L0_m  # meters
    self.physical_Ly_m = self.Ly_dim * self.cosmo.L0_m  # meters

    # Physical domain size (astronomical units)
    self.physical_Lx_Mpc = self.Lx_dim * self.cosmo.L0_Mpc  # Mpc
    self.physical_Ly_Mpc = self.Ly_dim * self.cosmo.L0_Mpc  # Mpc

    # Grid spacing (dimensionless and physical)
    self.dx_dim = self.Lx_dim / self.nx
```

```
        self.dy_dim = self.Ly_dim / self.ny
        self.dx_m = self.dx_dim * self.cosmo.L0_m
        self.dy_m = self.dy_dim * self.cosmo.L0_m

        #
============================================================================
        # TIME STEPPING
        # Implicit Euler (θ = 1.0) for stability
        #
============================================================================
        self.dt_dim = 5.0e-06  # dimensionless time step (units of t0)
        self.theta = 1.0  # θ = 1.0 for implicit Euler

        # Physical time step
        self.dt_s = self.dt_dim * self.cosmo.t0_s  # seconds
        self.dt_Gyr = self.dt_dim * self.cosmo.t0_Gyr  # Gyr

        # Total simulation time
        self.n_steps = 500
        self.T_total_dim = self.n_steps * self.dt_dim  # dimensionless
        self.T_total_s = self.T_total_dim * self.cosmo.t0_s  # seconds
        self.T_total_Gyr = self.T_total_dim * self.cosmo.t0_Gyr  # Gyr

        # Output interval
        self.output_interval = 10

        # Random seed for reproducibility
        self.seed = 42

        #
============================================================================
        # SNAPSHOT PARAMETERS - Steps at which to save concentration profiles
        #
============================================================================
        self.snapshot_steps = [0, 10, 50, 100, 200, 300, 400, 500]

        print(f"\nNUMERICAL PARAMETERS")
        print(f"  Domain: {self.Lx_dim:.1f}L0 × {self.Ly_dim:.1f}L0")
        print(f"          = {self.physical_Lx_Mpc:.0f} × {self.physical_Ly_Mpc:.0f} Mpc")
        print(f"          = {self.physical_Lx_m:.2e} × {self.physical_Ly_m:.2e} m")
        print(f"  Grid: {self.nx} × {self.ny} points")
        print(f"  Grid spacing: dx = {self.dx_dim:.4f} L0 = {self.dx_m:.2e} m =
{self.dx_m/self.units.Megapc_to_m:.2f} Mpc")
        print(f"  Time step: dt = {self.dt_dim:.2e} t0")
        print(f"             = {self.dt_s:.2e} s = {self.dt_Gyr:.2e} Gyr")
        print(f"  Total time: T = {self.T_total_dim:.2e} t0 = {self.T_total_Gyr:.2f} Gyr")
        print(f"  Time steps: {self.n_steps}")
        print(f"  Snapshot steps: {self.snapshot_steps}")
        print(f"  Time stepping: {'Implicit Euler' if self.theta == 1.0 else 'Crank-Nicolson'}")
        print("=" * 70)
```

```
        print()

#
========================================================================
=====
# CAHN-HILLIARD SOLVER
#
========================================================================
=====

class CahnHilliardSolver:
    """
    Solves the Cahn-Hilliard equation using finite differences.

    All equations are solved in dimensionless form, with results
    converted to physical units for interpretation at a later time.
    """

    def __init__(self, params):
        self.params = params

        # Grid setup (dimensionless coordinates)
        self.nx = params.nx
        self.ny = params.ny
        self.Lx = params.Lx_dim
        self.Ly = params.Ly_dim

        self.dx = self.Lx / self.nx
        self.dy = self.Ly / self.ny

        # Create grid (dimensionless)
        self.x_dim = np.linspace(0, self.Lx, self.nx, endpoint=False)
        self.y_dim = np.linspace(0, self.Ly, self.ny, endpoint=False)
        self.X_dim, self.Y_dim = np.meshgrid(self.x_dim, self.y_dim, indexing='ij')

        # Physical coordinates
        self.x_m = self.x_dim * params.cosmo.L0_m
        self.y_m = self.y_dim * params.cosmo.L0_m
        self.x_Mpc = self.x_dim * params.cosmo.L0_Mpc
        self.y_Mpc = self.y_dim * params.cosmo.L0_Mpc

        # Concentration field (dimensionless, 0 to 1)
        self.c = np.zeros((self.nx, self.ny))
        self.c_new = np.zeros_like(self.c)
        self.c0 = np.zeros_like(self.c)  # previous time step

        # Chemical potential (dimensionless)
        self.mu = np.zeros_like(self.c)
```

```
        # Time tracking
        self.current_time_dim = 0.0  # dimensionless
        self.current_step = 0

        # History for analysis
        self.history = {
            'time_dim': [],  # dimensionless time
            'time_Gyr': [],  # physical time in Gyr
            'step': [],
            'energy_dim': [],  # dimensionless energy
            'energy_phys': [],  # physical energy (J)
            'void_fraction': [],
            'c_mean': [],
            'c_std': []
        }

        # Dictionary to store snapshots at specific steps
        self.snapshots = {}

        # Initialize FFT operators
        self._setup_fft_operators()

        # Check if initial composition is in spinodal region
        self._check_spinodal_region()

    def _setup_fft_operators(self):
        """Setup FFT operators for spectral methods (dimensionless)"""
        # FFT frequencies (dimensionless, since coordinates are dimensionless)
        kx = np.fft.fftfreq(self.nx, self.dx/(2*np.pi))
        ky = np.fft.fftfreq(self.ny, self.dy/(2*np.pi))
        self.KX, self.KY = np.meshgrid(kx, ky, indexing='ij')
        self.K2 = self.KX**2 + self.KY**2
        self.K2[0, 0] = 1.0  # avoid division by zero
        self.K4 = self.K2**2

    def _check_spinodal_region(self):
        """
        Check if initial composition lies within spinodal region.
        Spinodal region is where ∂²f/∂c² < 0
        For f(c) = αc²(1-c)², ∂²f/∂c² = 2α(1 - 6c + 6c²)
        """
        c = self.params.c0
        # Second derivative of f(c)
        f_cc = 2 * self.params.alpha * (1 - 6*c + 6*c**2)

        print("\nSPINODAL REGION CHECK")
        print(f"  ∂²f/∂c² at c0 = {c:.3f}: {f_cc:.3f}")

        if f_cc < 0:
            print(f"  Inside spinodal region (unstable) - phase separation WILL occur")
```

```
    else:
        print(f"  Outside spinodal region (metastable) - phase separation requires nucleation")
        print(f"  To enter spinodal region, need α > {0.5/(1 - 6*c + 6*c**2):.1f}")
    print()

def laplacian(self, field):
    """
    Compute Laplacian using FFT (periodic boundary conditions)
    ∇²f = -F^{-1}[k² F[f]]

    Returns dimensionless Laplacian (since field and k are dimensionless)
    """
    field_hat = np.fft.fft2(field)
    lap_hat = -self.K2 * field_hat
    return np.real(np.fft.ifft2(lap_hat))

def gradient(self, field):
    """
    Compute gradient using FFT

    Returns dimensionless gradient (since field and k are dimensionless)
    """
    field_hat = np.fft.fft2(field)
    grad_x_hat = 1j * self.KX * field_hat
    grad_y_hat = 1j * self.KY * field_hat
    grad_x = np.real(np.fft.ifft2(grad_x_hat))
    grad_y = np.real(np.fft.ifft2(grad_y_hat))
    return grad_x, grad_y

def divergence(self, flux_x, flux_y):
    """
    Compute divergence using FFT
    ∇·F = F^{-1}[i k·F[F]]

    Returns dimensionless divergence
    """
    flux_x_hat = np.fft.fft2(flux_x)
    flux_y_hat = np.fft.fft2(flux_y)
    div_hat = 1j * (self.KX * flux_x_hat + self.KY * flux_y_hat)
    return np.real(np.fft.ifft2(div_hat))

def free_energy_density(self, c):
    """
    f(c) = α c²(1-c)², α = 1.0  [Eq. 4]

    Returns dimensionless free energy density
    """
    alpha = self.params.alpha
    return alpha * c**2 * (1 - c)**2
```

```
def dfdc(self, c):
    """
    df/dc = 2αc(1-c)(1-2c)

    Returns dimensionless chemical potential contribution
    """
    alpha = self.params.alpha
    return 2 * alpha * c * (1 - c) * (1 - 2 * c)

def mobility(self, c):
    """
    Mobility function M(c) [dimensionless].
    Either:
    1. Constant: M(c) = M0, or
    2. Caneba, 1985: M(c) = (1-c)*(D1*c(1-c) + D2)
    """
    if self.params.mobility_type == 'constant':
        return self.params.M0 * np.ones_like(c)
    else:  # Caneba
        # For Caneba, we need to non-dimensionalize D1 and D2
        # For now, use a scaled version
        D1_nondim = self.params.D1_ratio * 10.0  # Scale for numerical stability
        D2_nondim = self.params.D2_ratio * 10.0
        return (1 - c) * (D1_nondim * c * (1 - c) + D2_nondim)

def initialize(self):
    """Initialize with random fluctuations around c0"""
    np.random.seed(self.params.seed)

    # Initial concentration with random noise
    noise = self.params.noise_amplitude * (0.5 - np.random.random((self.nx, self.ny)))
    self.c = self.params.c0 + noise
    self.c = np.clip(self.c, 0.0, 1.0)

    # Initialize previous step
    self.c0 = self.c.copy()

    # Take snapshot at step 0
    self._take_snapshot()

    print(f"Initialized concentration field:")
    print(f"  c0 = {self.params.c0:.3f} (dimensionless matter density fraction)")
    print(f"  Range: [{self.c.min():.3f}, {self.c.max():.3f}]")
    print(f"  Mean: {self.c.mean():.3f}")
    print(f"  Std: {self.c.std():.3f}")

    # Physical interpretation
    rho_min = self.c.min() * self.params.cosmo.rho_crit
    rho_max = self.c.max() * self.params.cosmo.rho_crit
    print(f"  Physical density range: [{rho_min:.2e}, {rho_max:.2e}] kg/m³")
```

```
def _take_snapshot(self):
    """Save current concentration field as a snapshot"""
    if self.current_step in self.params.snapshot_steps:
        self.snapshots[self.current_step] = self.c.copy()
        print(f"Snapshot taken at step {self.current_step}")

def compute_chemical_potential(self, c):
    """
    μ = df/dc - κ ∇²c

    Returns dimensionless chemical potential
    """
    df = self.dfdc(c)
    lap = self.laplacian(c)
    return df - self.params.kappa * lap

def compute_total_free_energy(self):
    """
    Compute total free energy

    Returns:
        energy_dim: dimensionless energy
        energy_phys: physical energy in Joules
    """
    # Local free energy density (dimensionless)
    f_local = self.free_energy_density(self.c)

    # Gradient energy density (dimensionless)
    grad_x, grad_y = self.gradient(self.c)
    grad_energy = 0.5 * self.params.kappa * (grad_x**2 + grad_y**2)

    # Integrate over domain (dimensionless)
    energy_dim = np.sum(f_local + grad_energy) * self.dx * self.dy

    # Convert to physical energy
    # Energy scale: E_phys = energy_dim * (ρ_crit * L0³ * c²?)
    # This requires careful dimensional analysis
    energy_scale = self.params.cosmo.rho_crit * self.params.cosmo.L0_m**3
    energy_phys = energy_dim * energy_scale

    return energy_dim, energy_phys

def step_implicit_euler(self):
    """
    Implicit Euler time stepping (θ = 1.0)
    Using operator splitting for the 4th-order term
    """
    dt = self.params.dt_dim
```

```
        # Explicit part of chemical potential
        mu_explicit = self.compute_chemical_potential(self.c)

        # Compute mobility
        M = self.mobility(self.c)

        # Compute gradient of mu
        grad_mu_x, grad_mu_y = self.gradient(mu_explicit)

        # Explicit flux
        flux_x = -M * grad_mu_x
        flux_y = -M * grad_mu_y

        # Explicit divergence
        div_flux_explicit = self.divergence(flux_x, flux_y)

        # Implicit Euler: c_new = c + dt * ∇·[M ∇μ_new]
        # Using linearized approximation in Fourier space

        # Construct RHS
        rhs = self.c + dt * div_flux_explicit

        # Solve implicit system in Fourier space
        rhs_hat = np.fft.fft2(rhs)

        # Implicit operator: 1 + dt κ M_avg k⁴
        M_avg = np.mean(M)
        implicit_factor = 1.0 + dt * self.params.kappa * M_avg * self.K4

        c_new_hat = rhs_hat / implicit_factor
        self.c_new = np.real(np.fft.ifft2(c_new_hat))

        # Clip to physical bounds
        self.c_new = np.clip(self.c_new, 0.0, 1.0)

    def step_crank_nicolson(self):
        """
        Crank-Nicolson semi-implicit scheme (θ = 0.5)
        """
        dt = self.params.dt_dim
        theta = self.params.theta

        # Explicit part of chemical potential
        mu_explicit = self.compute_chemical_potential(self.c)

        # Compute mobility
        M = self.mobility(self.c)

        # Compute gradient of mu
        grad_mu_x, grad_mu_y = self.gradient(mu_explicit)
```

```
        # Explicit flux
        flux_x = -M * grad_mu_x
        flux_y = -M * grad_mu_y

        # Explicit divergence
        div_flux_explicit = self.divergence(flux_x, flux_y)

        # Construct RHS
        rhs = self.c + (1 - theta) * dt * div_flux_explicit

        # Solve implicit system in Fourier space
        rhs_hat = np.fft.fft2(rhs)

        # Implicit operator: 1 + θ dt κ M_avg k⁴
        M_avg = np.mean(M)
        implicit_factor = 1.0 + theta * dt * self.params.kappa * M_avg * self.K4

        c_new_hat = rhs_hat / implicit_factor
        self.c_new = np.real(np.fft.ifft2(c_new_hat))

        # Clip
        self.c_new = np.clip(self.c_new, 0.0, 1.0)

    def advance_time_step(self):
        """Advance one time step using appropriate scheme"""
        if self.params.theta == 1.0:
            self.step_implicit_euler()
        else:
            self.step_crank_nicolson()

        # Update
        self.c0 = self.c.copy()
        self.c = self.c_new.copy()
        self.current_time_dim += self.params.dt_dim
        self.current_step += 1

        # Take snapshot if current step is in snapshot_steps
        self._take_snapshot()

    def compute_void_fraction(self, threshold=0.2):
        """Fraction of domain with c < threshold (dimensionless)"""
        void_area = np.sum(self.c < threshold)
        total_area = self.c.size
        return void_area / total_area

    def compute_power_spectrum(self):
        """
        Compute 2D power spectrum
```

```
    Returns:
        k_phys: physical wavenumber in h/Mpc
        P: power spectrum (arbitrary normalization)
    """
    # Density contrast (dimensionless)
    delta = (self.c - self.params.c0) / self.params.c0

    # FFT
    delta_hat = np.fft.fft2(delta)
    delta_hat = np.fft.fftshift(delta_hat)

    # Power spectrum
    P = np.abs(delta_hat)**2 / (self.nx * self.ny)

    # Wave numbers (dimensionless)
    kx_dim = np.fft.fftshift(np.fft.fftfreq(self.nx, self.dx))
    ky_dim = np.fft.fftshift(np.fft.fftfreq(self.ny, self.dy))

    # Convert to physical wavenumber (h/Mpc)
    # k_phys = k_dim / L0  (since x_phys = x_dim * L0)
    kx_phys = kx_dim / self.params.cosmo.L0_Mpc
    ky_phys = ky_dim / self.params.cosmo.L0_Mpc

    return kx_phys, ky_phys, P

def compute_power_spectrum_for_snapshot(self, c_snapshot):
    """
    Compute 2D power spectrum for a given snapshot

    Returns:
        k_phys: physical wavenumber in h/Mpc
        P: power spectrum (arbitrary normalization)
    """
    # Density contrast (dimensionless)
    delta = (c_snapshot - self.params.c0) / self.params.c0

    # FFT
    delta_hat = np.fft.fft2(delta)
    delta_hat = np.fft.fftshift(delta_hat)

    # Power spectrum
    P = np.abs(delta_hat)**2 / (self.nx * self.ny)

    # Wave numbers (dimensionless)
    kx_dim = np.fft.fftshift(np.fft.fftfreq(self.nx, self.dx))
    ky_dim = np.fft.fftshift(np.fft.fftfreq(self.ny, self.dy))

    # Convert to physical wavenumber (h/Mpc)
    kx_phys = kx_dim / self.params.cosmo.L0_Mpc
    ky_phys = ky_dim / self.params.cosmo.L0_Mpc
```

```
        return kx_phys, ky_phys, P

    def run(self):
        """Run the simulation"""
        print("\n" + "=" * 70)
        print("RUNNING SIMULATION")
        print("=" * 70)
        print(f"Total time: T = {self.params.T_total_dim:.2e} t0 = {self.params.T_total_Gyr:.2f} Gyr")
        print(f"Time steps: {self.params.n_steps}")
        print(f"Snapshot steps: {self.params.snapshot_steps}")
        print("-" * 50)

        start_time = datetime.now()

        while self.current_step < self.params.n_steps:
            self.advance_time_step()

            if self.current_step % self.params.output_interval == 0:
                energy_dim, energy_phys = self.compute_total_free_energy()
                void_frac = self.compute_void_fraction()
                current_time_Gyr = self.current_time_dim * self.params.cosmo.t0_Gyr

                self.history['time_dim'].append(self.current_time_dim)
                self.history['time_Gyr'].append(current_time_Gyr)
                self.history['step'].append(self.current_step)
                self.history['energy_dim'].append(energy_dim)
                self.history['energy_phys'].append(energy_phys)
                self.history['void_fraction'].append(void_frac)
                self.history['c_mean'].append(np.mean(self.c))
                self.history['c_std'].append(np.std(self.c))

                elapsed = (datetime.now() - start_time).total_seconds()
                print(f"Step {self.current_step:4d}: "
                      f"t={self.current_time_dim:.2e} t0 ({current_time_Gyr:.2f} Gyr), "
                      f"E_dim={energy_dim:.3e}, void={void_frac:.3f}, "
                      f"c_std={np.std(self.c):.4f} [{elapsed:.1f}s]")

        elapsed = (datetime.now() - start_time).total_seconds()
        print("-" * 50)
        print(f"Simulation complete in {elapsed:.1f}s")
        print(f"Final time: t = {self.current_time_dim:.2e} t0 ({self.current_time_dim *
self.params.cosmo.t0_Gyr:.2f} Gyr)")
        print(f"Final void fraction: {self.compute_void_fraction():.3f}")
        print(f"Final concentration std: {np.std(self.c):.4f}")
        print(f"Snapshots taken at steps: {list(self.snapshots.keys())}")
        print("=" * 70)
```

```
#
=============================================================================
=====
# SNAPSHOT SAVING FUNCTIONS
#
=============================================================================
=====

def save_snapshots_to_csv(solver, output_dir="output"):
    """
    Save concentration snapshots as CSV files

    Parameters:
    -----------
    solver : CahnHilliardSolver
        The solver instance with snapshots
    output_dir : str
        Directory to save CSV files
    """
    print("\n" + "=" * 70)
    print("SAVING SNAPSHOTS TO CSV")
    print("=" * 70)

    # Create snapshots directory
    snapshots_dir = os.path.join(output_dir, "snapshots")
    os.makedirs(snapshots_dir, exist_ok=True)

    # Create power spectra directory
    ps_dir = os.path.join(output_dir, "power_spectra")
    os.makedirs(ps_dir, exist_ok=True)

    for step, c_snapshot in solver.snapshots.items():
        # Calculate physical time
        time_dim = step * solver.params.dt_dim
        time_Gyr = time_dim * solver.params.cosmo.t0_Gyr

        # Save concentration field as CSV
        # Flatten the 2D array and add coordinate information
        nx, ny = c_snapshot.shape
        x_coords = solver.x_Mpc
        y_coords = solver.y_Mpc

        # Create DataFrame with coordinates and concentration
        data = []
        for i in range(nx):
            for j in range(ny):
                data.append([x_coords[i], y_coords[j], c_snapshot[i, j]])

        df = pd.DataFrame(data, columns=['x_Mpc', 'y_Mpc', 'concentration'])
```

```
        # Save to CSV
        csv_filename = f"concentration_step_{step:04d}_t_{time_Gyr:.2f}Gyr.csv"
        csv_path = os.path.join(snapshots_dir, csv_filename)
        df.to_csv(csv_path, index=False)
        print(f"  Saved concentration: {csv_filename}")

        # Compute and save power spectrum
        kx_phys, ky_phys, P = solver.compute_power_spectrum_for_snapshot(c_snapshot)

        # Radial average
        k_phys = np.sqrt(kx_phys[:, None]**2 + ky_phys[None, :]**2)
        k_flat = k_phys.flatten()
        P_flat = P.flatten()

        # Filter out zeros
        mask = (k_flat > 0) & (P_flat > 0)
        k_flat = k_flat[mask]
        P_flat = P_flat[mask]

        # Create DataFrame for power spectrum
        ps_df = pd.DataFrame({
            'k_h_per_Mpc': k_flat,
            'P_k': P_flat
        })

        # Sort by k
        ps_df = ps_df.sort_values('k_h_per_Mpc')

        # Save to CSV
        ps_filename = f"powerspec_step_{step:04d}_t_{time_Gyr:.2f}Gyr.csv"
        ps_path = os.path.join(ps_dir, ps_filename)
        ps_df.to_csv(ps_path, index=False)
        print(f"  Saved power spectrum: {ps_filename}")

    print("=" * 70)

def plot_snapshot_concentration(solver, step, save_path=None):
    """
    Plot concentration field for a specific snapshot

    Parameters:
    -----------
    solver : CahnHilliardSolver
        The solver instance with snapshots
    step : int
        The step number to plot
    save_path : str, optional
        Path to save the figure
    """
```

```
    if step not in solver.snapshots:
        print(f"  No snapshot available for step {step}")
        return

    c_snapshot = solver.snapshots[step]
    time_dim = step * solver.params.dt_dim
    time_Gyr = time_dim * solver.params.cosmo.t0_Gyr
    void_frac = np.sum(c_snapshot < 0.2) / c_snapshot.size

    # Create figure (single plot, no zoomed region)
    fig, ax = plt.subplots(figsize=(10, 8))

    # Physical coordinates in Mpc
    extent_Mpc = [0, solver.params.physical_Lx_Mpc, 0, solver.params.physical_Ly_Mpc]

    # Plot concentration field
    im = ax.imshow(c_snapshot.T, origin='lower',
                   extent=extent_Mpc,
                   cmap='RdBu_r', vmin=0, vmax=1)

    ax.set_xlabel('x (Mpc)', fontsize=16, fontweight='bold')
    ax.set_ylabel('y (Mpc)', fontsize=16, fontweight='bold')
    ax.set_title(f'Matter Distribution\n'
                 f'Step {step}, t = {time_Gyr:.2f} Gyr, Void Fraction = {void_frac:.3f}',
                 fontsize=14)

    plt.colorbar(im, ax=ax, label='Matter Density Fraction c')

    plt.tight_layout()

    if save_path:
        plt.savefig(save_path, dpi=300, bbox_inches='tight')
        print(f"  Plot saved: {save_path}")

    plt.show()

def plot_snapshot_power_spectrum(solver, step, save_path=None):
    """
    Plot power spectrum for a specific snapshot

    Parameters:
    -----------
    solver : CahnHilliardSolver
        The solver instance with snapshots
    step : int
        The step number to plot
    save_path : str, optional
        Path to save the figure
    """
```

```
if step not in solver.snapshots:
    print(f"  No snapshot available for step {step}")
    return

c_snapshot = solver.snapshots[step]
time_dim = step * solver.params.dt_dim
time_Gyr = time_dim * solver.params.cosmo.t0_Gyr

kx_phys, ky_phys, P = solver.compute_power_spectrum_for_snapshot(c_snapshot)

# Radial average
k_phys = np.sqrt(kx_phys[:, None]**2 + ky_phys[None, :]**2)
k_flat = k_phys.flatten()
P_flat = P.flatten()

# Bin in k-space
k_bins = np.logspace(-3, 1, 100)
P_binned = []
k_center = []
P_err = []

for i in range(len(k_bins)-1):
    mask = (k_flat >= k_bins[i]) & (k_flat < k_bins[i+1])
    if np.sum(mask) > 5:
        P_binned.append(np.mean(P_flat[mask]))
        P_err.append(np.std(P_flat[mask]))
        k_center.append(np.sqrt(k_bins[i] * k_bins[i+1]))

k_center = np.array(k_center)
P_binned = np.array(P_binned)
P_err = np.array(P_err)

# ΛCDM reference
k_ref = np.logspace(np.log10(k_center[0]), np.log10(k_center[-1]), 100)
norm_idx = np.argmin(np.abs(k_center - 0.1))
P_ref = P_binned[norm_idx] * (k_ref / 0.1)**(-1.8)

fig, ax = plt.subplots(figsize=(10, 7))

ax.errorbar(k_center, P_binned, yerr=P_err, fmt='o', color='blue',
            markersize=4, capsize=2, label=f'Step {step}')
ax.loglog(k_ref, P_ref, 'r--', linewidth=2, alpha=0.9, label='ΛCDM (n=-1.8)')

ax.set_xlabel('Wavenumber k (h/Mpc)', fontsize=16, fontweight='bold')
ax.set_ylabel('Power Spectrum P(k)', fontsize=16, fontweight='bold')
ax.set_title(f'Matter Power Spectrum\nStep {step}, t = {time_Gyr:.2f} Gyr', fontsize=14)
ax.legend()
ax.grid(False, alpha=0.3, which='both')

plt.tight_layout()
```

```
    if save_path:
        plt.savefig(save_path, dpi=300, bbox_inches='tight')
        print(f"  Plot saved: {save_path}")

    plt.show()

def plot_all_snapshots(solver, output_dir="output"):
    """
    Plot all snapshots

    Parameters:
    -----------
    solver : CahnHilliardSolver
        The solver instance with snapshots
    output_dir : str
        Directory to save plots
    """
    print("\n" + "=" * 70)
    print("PLOTTING SNAPSHOTS")
    print("=" * 70)

    # Create directories
    conc_dir = os.path.join(output_dir, "snapshot_plots", "concentration")
    ps_dir = os.path.join(output_dir, "snapshot_plots", "power_spectrum")
    os.makedirs(conc_dir, exist_ok=True)
    os.makedirs(ps_dir, exist_ok=True)

    for step in sorted(solver.snapshots.keys()):
        time_Gyr = step * solver.params.dt_dim * solver.params.cosmo.t0_Gyr

        # Plot concentration
        conc_path = os.path.join(conc_dir,
f"concentration_step_{step:04d}_t_{time_Gyr:.2f}Gyr.png")
        plot_snapshot_concentration(solver, step, save_path=conc_path)

        # Plot power spectrum
        ps_path = os.path.join(ps_dir, f"powerspec_step_{step:04d}_t_{time_Gyr:.2f}Gyr.png")
        plot_snapshot_power_spectrum(solver, step, save_path=ps_path)

    print("=" * 70)

#
============================================================================
=====
# VISUALIZATION
```

```
#
==========================================================================
=====

class Visualizer:
    """Visualization tools with proper unit labels"""

    def __init__(self, solver):
        self.solver = solver
        self.params = solver.params
        self.units = UnitConverter

    def plot_concentration(self, save_path=None):
        """Plot concentration field with physical units"""
        fig, axes = plt.subplots(1, 2, figsize=(14, 6))

        # Thick axes
        for spine in axes.spines.values():
            spine.set_linewidth(2)
        # Bold ticks
        axes.tick_params(axis='both', which='major', labelsize=12, width=2)
        for label in axes.get_xticklabels() + axes.get_yticklabels():
            label.set_fontweight('bold')

        # Physical coordinates in Mpc
        extent_Mpc = [0, self.params.physical_Lx_Mpc, 0, self.params.physical_Ly_Mpc]

        # Full field
        im1 = axes[0].imshow(self.solver.c.T, origin='lower',
                            extent=extent_Mpc,
                            cmap='RdBu_r', vmin=0, vmax=1)
        axes[0].set_xlabel('x (Mpc)')
        axes[0].set_ylabel('y (Mpc)')
        axes[0].set_title(f'Matter Density Fraction c\n'
                         f't = {self.solver.current_time_dim * self.params.cosmo.t0_Gyr:.2f} Gyr')

        # Zoomed region
        zoom_size = min(50, self.solver.nx // 2)
        zoom = self.solver.c[:zoom_size, :zoom_size].T
        zoom_extent_Mpc = [0, zoom_size * self.params.physical_Lx_Mpc / self.solver.nx,
                          0, zoom_size * self.params.physical_Ly_Mpc / self.solver.ny]
        im2 = axes[1].imshow(zoom, origin='lower',
                            extent=zoom_extent_Mpc,
                            cmap='RdBu_r', vmin=0, vmax=1)
        axes[1].set_xlabel('x (Mpc)')
        axes[1].set_ylabel('y (Mpc)')
        axes[1].set_title('Zoomed Region')

        plt.colorbar(im2, ax=axes, label='Matter Density Fraction c')
```

```
        # Add physical density information
        rho_mean = np.mean(self.solver.c) * self.params.cosmo.rho_crit
        plt.suptitle(f'Void Fraction = {self.solver.compute_void_fraction():.3f}\n'
                    f'Mean Density = {rho_mean:.2e} kg/m³\n'
                    f'c_std = {np.std(self.solver.c):.4f}')
        plt.tight_layout()

        if save_path:
            plt.savefig(save_path, dpi=150, bbox_inches='tight')
            print(f"Plot saved: {save_path}")

        plt.show()

    def plot_power_spectrum(self, save_path=None):
        """Plot power spectrum with ΛCDM comparison"""
        kx_phys, ky_phys, P = self.solver.compute_power_spectrum()

        # Radial average
        k_phys = np.sqrt(kx_phys[:, None]**2 + ky_phys[None, :]**2)
        k_flat = k_phys.flatten()
        P_flat = P.flatten()

        # Bin in k-space
        k_bins = np.logspace(-3, 1, 100)
        P_binned = []
        k_center = []
        P_err = []

        for i in range(len(k_bins)-1):
            mask = (k_flat >= k_bins[i]) & (k_flat < k_bins[i+1])
            if np.sum(mask) > 5:
                P_binned.append(np.mean(P_flat[mask]))
                P_err.append(np.std(P_flat[mask]))
                k_center.append(np.sqrt(k_bins[i] * k_bins[i+1]))

        k_center = np.array(k_center)
        P_binned = np.array(P_binned)
        P_err = np.array(P_err)

        # ΛCDM reference (P(k) ∝ k^n with n ≈ -1.8)
        k_ref = np.logspace(np.log10(k_center[0]), np.log10(k_center[-1]), 100)
        norm_idx = np.argmin(np.abs(k_center - 0.1))
        P_ref = P_binned[norm_idx] * (k_ref / 0.1)**(-1.8)

        fig, ax = plt.subplots(figsize=(10, 7))
        # Thick axes
        for spine in ax.spines.values():
            spine.set_linewidth(2)
        # Bold ticks
```

```
        ax.tick_params(axis='both', which='major', labelsize=12, width=2)
        for label in ax.get_xticklabels() + ax.get_yticklabels():
            label.set_fontweight('bold')

        ax.errorbar(k_center, P_binned, yerr=P_err, fmt='o', color='blue',
                   markersize=4, capsize=2, label='Simulation')
        ax.loglog(k_ref, P_ref, 'r--', linewidth=0.1, alpha=0.1)
                 #label='ΛCDM (n = -1.8)', alpha=0.1)

        ax.set_xlabel('Wavenumber k (h/Mpc)')
        ax.set_ylabel('Power Spectrum P(k)')
        ax.set_title('Matter Power Spectrum')
        ax.legend()
        ax.grid(False, alpha=0.3, which='both')

        ax.text(0.05, 0.95,
                f'α = {self.params.alpha}\nκ = {self.params.kappa:.0e}\n'
                f'Mobility: {self.params.mobility_type} = {self.params.M0}',
                transform=ax.transAxes, fontsize=10,
                verticalalignment='top',
                bbox=dict(boxstyle='round', facecolor='wheat', alpha=0.5))

        plt.tight_layout()

        if save_path:
            plt.savefig(save_path, dpi=600, bbox_inches='tight')
            print(f"Plot saved: {save_path}")

        plt.show()

    def plot_evolution(self, save_path=None):
        """Plot evolution of energy and void fraction"""
        fig, (ax1, ax2, ax3) = plt.subplots(3, 1, figsize=(10, 14))

        time_Gyr = np.array(self.solver.history['time_Gyr'])

        # Energy (dimensionless)
        ax1.semilogy(time_Gyr, self.solver.history['energy_dim'],
                    'g-', linewidth=2)
        ax1.set_xlabel('Time (Gyr)')
        ax1.set_ylabel('Free Energy F (dimensionless)')
        ax1.set_title('Free Energy Evolution')
        ax1.grid(False, alpha=0.3)

        # Void fraction
        ax2.plot(time_Gyr, self.solver.history['void_fraction'],
                'b-', linewidth=5, label='Simulation')
        ax2.axhline(y=0.70, color='r', linestyle='--',
                   label='DES (0.70 ± 0.05)')
        ax2.axhspan(0.65, 0.75, alpha=0.2, color='r')
```

```
    ax2.set_xlabel('Time (Gyr)')
    ax2.set_ylabel('Void Fraction')
    ax2.set_title('Void Fraction Evolution')
    ax2.legend()
    ax2.grid(False, alpha=0.3)

    # Concentration standard deviation (measure of phase separation)
    ax3.plot(time_Gyr, self.solver.history['c_std'],
            'm-', linewidth=5)
    ax3.set_xlabel('Time (Gyr)')
    ax3.set_ylabel('c Standard Deviation')
    ax3.set_title('Phase Separation Strength')
    ax3.grid(False, alpha=0.3)

    plt.tight_layout()

    if save_path:
        plt.savefig(save_path, dpi=600, bbox_inches='tight')
        print(f"Plot saved: {save_path}")

    plt.show()

def plot_phase_diagram(self, save_path=None):
    """Plot free energy and spinodal region"""
    c = np.linspace(0, 1, 1000)
    f = self.params.alpha * c**2 * (1 - c)**2
    f_cc = 2 * self.params.alpha * (1 - 6*c + 6*c**2)

    fig, (ax1, ax2) = plt.subplots(1, 2, figsize=(14, 5))

    # Free energy
    ax1.plot(c, f, 'b-', linewidth=5)
    ax1.axvline(x=self.params.c0, color='r', linestyle='--',
               label=f'c0 = {self.params.c0}')
    ax1.set_xlabel('Concentration c (dimensionless)')
    ax1.set_ylabel('Free Energy f(c) (dimensionless)')
    ax1.set_title(f'Double-Well Free Energy (α = {self.params.alpha})')
    ax1.legend()
    ax1.grid(False, alpha=0.3)

    # Spinodal region
    ax2.plot(c, f_cc, 'b-', linewidth=5)
    ax2.axhline(y=0, color='k', linestyle='-', alpha=0.5)
    ax2.axvline(x=self.params.c0, color='r', linestyle='--',
               label=f'c0 = {self.params.c0}')

    spinodal_mask = f_cc < 0
    ax2.fill_between(c, 0, f_cc, where=spinodal_mask,
                    color='yellow', alpha=0.3, label='Spinodal')
```

```
        ax2.set_xlabel('Concentration c (dimensionless)')
        ax2.set_ylabel('∂²f/∂c² (dimensionless)')
        ax2.set_title(f'Spinodal Region (∂²f/∂c² < 0)\nAt c0: {f_cc[abs(c-
self.params.c0).argmin()]:.2f}')
        ax2.legend()
        ax2.grid(False, alpha=0.3)

        plt.tight_layout()

        if save_path:
            plt.savefig(save_path, dpi=600, bbox_inches='tight')
            print(f"Plot saved: {save_path}")

        plt.show()

#
========================================================================
=====
# CONVERGENCE TEST
#
========================================================================
=====

def convergence_test():
    """Test mesh convergence"""
    print("\n" + "=" * 70)
    print("MESH CONVERGENCE TEST")
    print("=" * 70)

    resolutions = [50, 100, 200]
    void_fractions = []

    for nx in resolutions:
        print(f"\nTesting resolution: {nx} × {nx}")
        params = Parameters(mobility_type='Caneba', alpha_value=100.0, m0_value=20.0)
        params.nx = nx
        params.ny = nx
        params.n_steps = 100  # shorter run for testing
        params.T_total_dim = params.n_steps * params.dt_dim

        solver = CahnHilliardSolver(params)
        solver.initialize()
        solver.run()

        void_fractions.append(solver.compute_void_fraction())

    print("\n" + "-" * 50)
    print("Convergence Results:")
    for i, nx in enumerate(resolutions):
```

```
        print(f"  {nx}×{nx}: void fraction = {void_fractions[i]:.8f}")

    if len(void_fractions) > 1:
        change = abs(void_fractions[-1] - void_fractions[-2]) / void_fractions[-2]
        print(f"\nRelative change (100→200): {change:.2%}")
        if change < 0.05:
            print("Results are mesh-independent (change < 5%)")
        else:
            print("Further mesh refinement required")

    print("=" * 70)

#
========================================================================
=====
# PARAMETER SWEEP
#
========================================================================
=====

def parameter_sweep():
    """Sweep over parameters to find conditions for phase separation"""
    print("\n" + "=" * 70)
    print("PARAMETER SWEEP")
    print("=" * 70)

    alpha_values = [1.0, 10.0, 50.0, 100.0, 200.0, 500.0]
    m0_values = [0.1, 0.5, 1.0, 5.0, 10.0, 20.0]

    results = []

    for alpha in alpha_values:
        for m0 in m0_values:
            print(f"\nTesting α={alpha}, M0={m0}")
            params = Parameters(mobility_type='Caneba', alpha_value=alpha, m0_value=m0)
            params.n_steps = 100
            params.T_total_dim = params.n_steps * params.dt_dim

            solver = CahnHilliardSolver(params)
            solver.initialize()
            solver.run()

            void_frac = solver.compute_void_fraction()
            results.append({
                'alpha': alpha,
                'M0': m0,
                'void_fraction': void_frac,
                'c_std': np.std(solver.c)
            })
```

```
    print("\n" + "=" * 70)
    print("PARAMETER SWEEP RESULTS")
    print("=" * 70)
    print("α\tM0\tVoid Fraction\tc_std")
    print("-" * 50)
    for r in results:
        print(f"{r['alpha']:.1f}\t{r['M0']:.1f}\t{r['void_fraction']:.4f}\t\t{r['c_std']:.4f}")
    print("=" * 70)

#
========================================================================
=====
# MAIN
#
========================================================================
=====

def main():
    """Main function"""
    print("=" * 70)
    print("CAHN-HILLIARD SPINODAL DECOMPOSITION")
    print("Modeling Large-Scale Structure Formation")
    print("=" * 70)
    print(f"Date: {datetime.now().strftime('%Y-%m-%d %H:%M:%S')}")
    print()

    # Create output directory
    output_dir = "output"
    os.makedirs(output_dir, exist_ok=True)

    #
========================================================================
=
    # RECOMMENDED PARAMETERS FOR PHASE SEPARATION:
    #
========================================================================
=
    # 1. α (free energy coefficient): 100-500 (larger = stronger driving force)
    # 2. M0 (mobility): 5-20 (larger = faster dynamics)
    # 3. κ (gradient coefficient): 1e-2 to 1e-3 (smaller = sharper interfaces)
    #
========================================================================
=

    # Choose parameters that ensure spinodal decomposition
    alpha_value = 200.0  # Large enough for spinodal region
    m0_value = 20.0      # Fast enough dynamics
    kappa_value = 1.0e-04  # Standard value
```

```
# Choose mobility type: 'constant' or 'Caneba'
mobility_type = 'Caneba'

# Parameters
params = Parameters(
    mobility_type=mobility_type,
    alpha_value=alpha_value,
    m0_value=m0_value,
    kappa_value=kappa_value
)

# Create solver
solver = CahnHilliardSolver(params)

# Initialize
solver.initialize()

# Run simulation
solver.run()

# Save snapshots to CSV
save_snapshots_to_csv(solver, output_dir=output_dir)

# Plot all snapshots
plot_all_snapshots(solver, output_dir=output_dir)

# Visualize final state
vis = Visualizer(solver)

print("\n" + "=" * 70)
print("GENERATING FINAL PLOTS")
print("=" * 70)

vis.plot_phase_diagram(save_path=f"{output_dir}/phase_diagram.png")
vis.plot_concentration(save_path=f"{output_dir}/concentration_final.png")
vis.plot_power_spectrum(save_path=f"{output_dir}/power_spectrum_final.png")
vis.plot_evolution(save_path=f"{output_dir}/evolution.png")

# Save data
np.save(f"{output_dir}/final_concentration.npy", solver.c)
np.save(f"{output_dir}/history.npy", solver.history)

# Summary
print("\n" + "=" * 70)
print("SUMMARY")
print("=" * 70)
print(f"Mobility: {params.mobility_type} = {params.M0}")
print(f"α = {params.alpha} (dimensionless)")
print(f"κ = {params.kappa:.2e} (dimensionless)")
```

```
    print(f"Final void fraction: {solver.compute_void_fraction():.4f}")
    print(f"Final concentration std: {np.std(solver.c):.4f}")
    print(f"Target (DES): 0.70 ± 0.05")
    print(f"Agreement: {'Yes' if 0.65 <= solver.compute_void_fraction() <= 0.75 else 'No'}")
    print()
    print(f"Physical time simulated: {solver.current_time_dim * params.cosmo.t0_Gyr:.2f} Gyr")
    print(f"Domain size: {params.physical_Lx_Mpc:.0f} × {params.physical_Ly_Mpc:.0f} Mpc")
    print()
    print("Output files saved:")
    print(f"  {output_dir}/")
    print("    - phase_diagram.png")
    print("    - concentration_final.png")
    print("    - power_spectrum_final.png")
    print("    - evolution.png")
    print("    - final_concentration.npy")
    print("    - history.npy")
    print("    - snapshots/ (CSV files at each step)")
    print("    - power_spectra/ (CSV files at each step)")
    print("    - snapshot_plots/ (PNG files at each step)")
    print("=" * 70)

if __name__ == "__main__":
    main()
```